\documentclass[9pt,reqno]{amsart}
\newtheorem{thm}{Theorem}[section]
\newtheorem{defi}{Definition}[section]

\newtheorem{cor}{Corollary}[section]
\newtheorem{pr}{Proposition}[section]
\theoremstyle{definition}
\newtheorem{rem}{Remark}[section]

\newcommand{\be}{\begin{equation}}
\newcommand{\ee}{\end{equation}}
\newcommand{\bea}{\begin{eqnarray}}
\newcommand{\eea}{\end{eqnarray}}
\newcommand{\beb}{\begin{eqnarray*}}
\newcommand{\eeb}{\end{eqnarray*}}
\usepackage{amssymb,amsfonts,amsthm,setspace,indentfirst,mathrsfs}
\usepackage{minibox}
\usepackage{enumitem}
\usepackage{hyperref}
\usepackage{cite}
\usepackage{color}
\numberwithin{equation}{section}
\begin{document}
%%%%%%%%%%%%%%%%%%%%%%%%%%%%%%%%%%%%%%%%%%%%%%%%%%%%%%%%%%%%%%%%%%%%%%%%%%%%%%%%%%%%%%%%%%%%%%%%%%%%%%%%%%%%%
%
\title[ Geometrical properties of a Point-Like Global Monopole Spacetime ]{\bf{ Geometrical properties of a Point-Like Global Monopole SpaceTime }}

\author[A. A. Shaikh,   F. Ahmed \& B. R. Datta]{Absos Ali Shaikh$^{1}$,  Faizuddin Ahmed$^{2}$ and Biswa Ranjan Datta$^{3}$}
\date{\today}

\address{\noindent$^{1,3}$ Department of Mathematics,
	\newline University of Burdwan, Golapbag,
	\newline Burdwan-713104, West Bengal, India}
\email{aashaikh@math.buruniv.ac.in$^{1}$ ; aask2003@yahoo.co.in$^{1}$ }
\email{biswaranjandatta2019@gmail.com$^{3}$}

\address{\noindent$^2$ Department of Physics, 
	\newline University of Science \& Technology Meghalaya, 
	\newline Ri-Bhoi, Meghalaya-793101, India,}
\email{faizuddinahmed15@gmail.com$^2$ ; faizuddin@ustm.ac.in$^2$ (corresponding author)}

\dedicatory{}
%%%%%%%%%%%%%%%%%%%%%%%%%%%%%%%%%%%%%%%%%%%%% Abstract %%%%%%%%%%%%%%%%%%%%%%%%%%%%%%%%%%%%%%%%%%%%%%%%%%%%%
\begin{abstract}
The aim of this paper is to study the geometric properties of the point-like global monopole (briefly, PGM) spacetime, which is a static and spherically symmetric solution of the Einstein's field equations. It has shown that PGM spacetime  admits various types of  pseudosymmetry structures, such as pseudosymmetry due to Weyl conformal curvature tensor,  pseudosymmetry due to concircular curvature tensor, pseudosymmetry due to conharmonic curvature tensor,  Ricci generalized conformal pseudo-symmetric due to projective curvature tensor, Ricci generalized projective pseudo-symmetric. Moreover,  it has proved that PGM spacetime is $2$-quasi Einstein, generalized quasi-Einstein,  Einstein manifold of degree $2$, and its Weyl conformal curvature $2$-forms are recurrent. The energy-momentum tensor of the PGM spacetime realizes several types of pseudosymmetry, and its Ricci tensor is compatible with Riemann curvature, Weyl conformal curvature, projective curvature, and conharmonic curvature and concircular curvature. Further, it has shown that PGM spacetime admits motion, curvature collineation, and Ricci collineation. Also, the notion of curvature inheritance (resp., curvature collineation) for the (1,3)-type curvature tensor is not equivalent to the notion of curvature inheritance (resp., curvature collineation) for the (0,4)-type curvature tensor as it has shown that such distinctive properties were possessed by PGM spacetime. Hence the notions of curvature inheritance defined by Duggal \cite{Duggal1992} and Shaikh and Datta \cite{ShaikhDatta2022} are not equivalent.
\end{abstract}
%%%%%%%%%%%%%%%%%%%%%%%%%%%%%%%%%%%%%%%%%%%%%%%%%%%%%%%%%%%%%%%%%%%%%%%%%%%%%%%%%%%%%%%%%%%%%%%%%%%%%%%%%%%%
%
\maketitle
%

%%%%%%%%%%%%%%%%%%%%%%%%%%%%%%%%%%%%%%%%%%%%%%%%%%%%%%%%%%%%%%%%%%%%%%%%%%%%%%%%%%%%%%%%%%%%%%%%%%%%%%%%%%%%%%
%																					Introduction
%%%%%%%%%%%%%%%%%%%%%%%%%%%%%%%%%%%%%%%%%%%%%%%%%%%%%%%%%%%%%%%%%%%%%%%%%%%%%%%%%%%%%%%%%%%%%%%%%%%%%%%%%%%%%%
%\section{\bf Introduction}\label{intro}

\section{Introduction}
\label{intro}

\indent Let  $M$ be a smooth and connected manifold of dimension $n$ $(\geq 3)$ equipped with a semi-Riemannian metric $g$ of signature $(\delta, n-\delta)$, $0\leq \delta\leq n$. If $\delta=1 \text{ or } n-1$ (resp.,  $\delta=0 \text{ or } n$), then $M$ is known as a Lorentzian (resp., Riemannian) manifold, and spacetimes are the mathematical models of $4$-dimensional connected Lorentzian manifolds. Throughout the paper $\nabla, R, S, \kappa$  respectively denote the Levi-Civita connection, Riemann curvature tensor of type $(0,4)$, Ricci tensor of type $(0,2)$, and the scalar curvature of $M$.

Understanding the symmetry of a semi-Riemannian manifold, curvature rigorous performs a crucial role as in the year 1926, Cartan introduced the notion of local symmetry  \cite{Cart26} by the relation $\nabla R=0$  and the notion of semisymmetry \cite{Cart46} by the relation $R \cdot R=0$ in 1946 (later classified by Szab\'{o} \cite{Szab82,Szab84,Szab85}). During the last eight decades, several authors debilitated such curvature conditions to generalize the concept of symmetry in various directions, which infers different generalized notions of symmetry, such as pseudosymmetric manifolds by Chaki \cite{Chak87, Chak88},  pseudosymmetric manifolds by Adam\'{o}w and Deszcz \cite{AD83}, weakly symmetric manifolds by Tam\'{a}ssy and Binh \cite{TB89, TB93}, recurrent   manifolds by Ruse \cite{Ruse46, Ruse49a, Ruse49b, Walk50}, curvature $2$-forms of recurrent manifolds  \cite{Bess87, LR89}, several kinds of generalized recurrent manifolds by  Shaikh \textit{et al.} \cite{SAR13, SP10, SR10, SR11, SRK15, SRK17} etc. We note that Som-Raychaudhuri spacetime \cite{SK16}, Robertson-Walker spacetime \cite{ADEHM14,DK99}, G\"{o}del spacetime \cite{DHJKS14}, Siklos spacetime \cite{SDKC19}, Robinson-Trautman spacetime \cite{SAA18} and  Reissner-Nordstr\"{o}m spacetime \cite{Kowa06} admit different pseudosymmetric type geometric structures.

There are two major aspects of geometric structures of a certain spacetime, one is for geometry and another is its physical nature due to the Einstein field equation (briefly, EFE). The main moto of this paper is to explore the geometric structures of PGM spacetime in terms of curvatures appearing by means of first order as well as higher order covariant derivatives.

Again, to constitute gravitational potentials satisfying EFE, imposing symmetry is a vital tool, which implies that the geometrical symmetries play a crucial role in the theory of general relativity. A geometric quantity is preserved along a vector field if the Lie derivative of a certain tensor vanishes concerning that vector field, and the vanishing Lie derivative illustrates geometrical symmetries. Motion, curvature collineation, Ricci collineation etc. are  the notions of such symmetries. Katzin \textit{et al.} \cite{KLD1969,KLD1970} rigorously investigated the role of curvature collineation in general relativity. In 1992, Duggal \cite{Duggal1992}  introduced the notion of curvature inheritance, generalizing the concept of curvature collineation for the (1,3)-type curvature tensor. Recently, Shaikh and Datta \cite{ShaikhDatta2022} introduced the concept of generalized curvature inheritance, that is, a generalization of curvature collineation and curvature inheritance for the (0,4)-type curvature tensor. During the last three decades, plenty of papers (see, for example, \cite{Ahsan1978,Ahsan1977_231,Ahsan1977_1055,Ahsan1987,Ahsan1995,Ahsan1996,Ahasan2005,AhsanAli2014,AA2012, 
AH1980,AliAhsan2012,SASZ2022,ShaikhDatta2022}) have reported in the literature regarding investigations of such kinds of symmetries. Here, we have shown that the PGM spacetime admits several symmetries, such as motion, curvature collineation, Ricci collineation and curvature inheritance. Also, it has shown  that the notions of curvature inheritance and curvature collineation for the (1,3)-type curvature tensor by Duggal \cite{Duggal1992} and for the (0,4)-type curvature tensor by Shaikh {\it et al.} \cite{ShaikhDatta2022} are not equivalent as PGM spacetime realizes such distinctive properties.

Some Grand Unified Theories suggested that the topological defects formed during the phase transition in the early universe through a spontaneous symmetry-breaking mechanism \cite{TWBK,AVV}. In literature, different kinds of topological defects have been known and among them, cosmic strings \cite{WAH} and global monopoles \cite{MBAV,ERFM2} have been widely investigated. A global monopole is a heavy object characterized by spherical symmetry and divergent mass. The gravitational field of a static global monopole has been discussed by Barriola {\it et al.} \cite{MBAV} and is expected to be stable against spherical and polar perturbation. This global monopole has been studied in the context of general relativity and quantum mechanical systems (see, for example, Refs. \cite{ae,ALCO2,ERBM,ERBM44,AAS,AAS2,AAS3,AAS4,AAS5,AAS6}).

In Ref. \cite{MBAV}, the authors discussed a point-like global monopole spacetime and its features in Ref. \cite{ERFM2}, where quantum mechanical problems have also been investigated. Therefore, line- element of the point-like global monopole (PGM) spacetime which is a static and spherically symmetric metric in the spherical coordinates $(t, r, , \theta,\phi)$ is described by \cite{MBAV,ERFM2}
\begin{equation}
ds^2=-dt^2+\frac{dr^2}{\alpha^2}+r^2\,(d\theta^2+\sin^2 \theta\,d\phi^2),
\label{1}
\end{equation}
where $\alpha^2=\big(1-8\pi\eta_0^2\big)<1$ depends on the energy scale $\eta_0$. The parameter $\eta_0$ represents the dimensionless volumetric mass density of the PGM defect. Here, the different coordinates are in the ranges $-\infty < t < +\infty$,\quad $0 \leq r < \infty$,\quad $0 \leq \theta \leq \frac{\pi}{2}$, and $0 \leq \phi < 2\,\pi$. The PGM spacetime reveals some interesting features delineated as: (i) it is not globally flat, and possesses a naked curvature singularity on the axis given by the Ricci scalar $\kappa=\frac{2\,(\alpha^2-1)}{r^2}$, (ii) the area of a sphere of radius $r$ in this manifold is not $4\,\pi\,r^2$ but rather it is equal to $4\,\pi\,\alpha^2\,r^2$, (iii) its surface $\theta=\frac{\pi}{2}$ presents the geometry of a cone with the deficit angle $\nabla\,\phi=8\,\pi^2\,\eta_0^2$, and (iv) there is no Newtonian-like gravitational potential: $g_{tt}=-1$. Other interesting features of this PGM spacetime have been given in details in \cite{ERFM2}. If $\alpha \to 1$, one can obtain the spherically symmetric Minkowski flat space line-element. It is better to mention here that the PGM spacetime has been studied by many authors in quantum system both in the relativistic limit \cite{ALCO,ABHA,EAFB,SZ,SR2,SZ3} as well as the non-relativistic limit \cite{CF44,RV,GAM,PN,FA,FA2,FA3,FA4,FA5,FA6,FA7,FA8,FA9,FA10}.

The present paper exhibits several curvature properties of PGM spacetime accomplished by the metric (\ref{1}), such as pseudosymmetry due to Weyl conformal curvature, pseudosymmetry due to concircular curvature, pseudosymmetry due to conharmonic curvature  etc.. Also, we show that this spacetime is neither Einstein nor quasi-Einstein, but an Einstein manifold of level $2$, generalized quasi Einstein, $2$-quasi-Einstein manifold and conformal curvature $2$-forms are recurrent. Moreover, it is shown that this metric is Ricci generalized conformal pseudosymmetric due to projective curvature tensor, Ricci generalized projective pseudosymmetric etc., also Ricci tensor is neither Codazzi type nor cyclic parallel. Additionally, we show that the energy-momentum tensor also admits several types of pseudosymmetric structures, and the Ricci tensor is compatible with Riemann, conformal, projective, concircular and conharmonic curvature tensors. Finally, it is shown that for certain vector fields, the PGM spacetime reveals motion, curvature collineation, Ricci collineation and curvature inheritance. Also, by exhibiting some distinctive properties of PGM spacetime, it is shown that the notions of curvature inheritance for (1,3)-type curvature tensor and the (0,4)-type curvature tensor are not equivalent.

The paper is organized as follows: In Sect. $2$, we discuss various rudimentary facts regarding curvature tensors with their derivatives and different pseudosymmetric structures, which are essential throughout this paper to investigate geometric properties of PGM spacetime. In Sect. $3$, we have computed several pseudosymmetric structures admitted by PGM spacetime and geometric structures with other features. In Sect. $4$, we have investigated geometric structures of such a spacetime due to the energy-momentum tensor. Sect. $5$ deals with  symmetries associated with PGM spacetime, such as motion, curvature collineation, Ricci collineation, and curvature inheritance. Also, the distinctness of the notions of curvature inheritance for $(1,3)$-type curvature tensor and for $(0,4)$-type curvature tensor is presented. Finally, the last section consists of the conclusion of the paper briefly.

\section{\bf Preliminaries}

This section aims to explain different kinds of geometric structures originated by appointing restrictions on the curvatures and their covariant derivatives, which are effective to elaborate the symmetry of the PGM spacetime having certain geometric meanings. Also, the notions of motion, curvature inheritance and Ricci inheritance are illustrated in this section.

For two symmetric second-order covariant tensors $\nu_1$ and $\nu_2$, the Kulkarni-Nomizu product $\nu_1\wedge \nu_2$
is defined by (see, \cite{DGHS11,Glog02,SKA18})
\begin{eqnarray*}
(\nu_1\wedge \nu_2)(\eta_1,\eta_2,\iota_1,\iota_2)&=&\nu_1(\eta_1,\iota_2)\nu_2(\eta_2,\iota_1) + \nu_1(\eta_2,\iota_1)\nu_2(\eta_1,\iota_2)\\ &-& \nu_1(\eta_1,\iota_1)\nu_2(\eta_2,\iota_2) - \nu_1(\eta_2,\iota_2)\nu_2(\eta_1,\iota_1),
\end{eqnarray*}
where $\eta_1, \eta_2, \iota_1,\iota_2$ $\in \chi(M),$ the Lie algebra of all smooth vector fields on $M$.
Now, We define some endomorphisms given as follows:
(see, \cite{DGHS11,DHJKS14,Glog02,Kowa06, SC21,SK19})
\begin{eqnarray*}
	(\eta_1 \wedge_{\nu} \eta_2)\iota &=& \nu(\eta_2,\iota)\eta_1-\nu(\eta_1,\iota)\eta_2,\\
	\mathcal{I}_R(\eta_1,\eta_2) &=& [\nabla_{\eta_1},\nabla_{\eta_2}] - \nabla_{[\eta_1,\eta_2]},\\
	\mathcal{I}_C(\eta_1,\eta_2) &=& \mathcal{I}_R(\eta_1,\eta_2) - \frac{1}{n-2}\\
								& &\times \left(\mathscr{E} \eta_1 \wedge_g \eta_2 +  \eta_1 \wedge_g \mathscr{E} \eta_2 - \frac{\kappa}{n-1}\eta_1 \wedge_g \eta_2\right),\\
	\mathcal{I}_K (\eta_1,\eta_2)&=& \mathcal{I}_R (\eta_1,\eta_2)- \frac{1}{n-2}\left(\mathscr{E} \eta_1 \wedge_g \eta_2 +  \eta_1 \wedge_g \mathscr{E} \eta_2\right),\\
	\mathcal{I}_W (\eta_1,\eta_2)&=& \mathcal{I}_R (\eta_1,\eta_2) - \frac{\kappa }{n (n-1)}\ \eta_1 \wedge_g \eta_2,\\
	\mathcal{I}_P (\eta_1,\eta_2)&=& \mathcal{I}_R(\eta_1,\eta_2) -\frac{1}{n-1}\ \eta_1 \wedge_S \eta_2,
	\end{eqnarray*}
where $\nu$ is a (0,2) type symmetric tensor and $\mathscr{E}$ is the Ricci operator  defined by $S(\eta_1,\eta_2)=g(\eta_1,\mathscr{E}(\eta_2))$. Throughout this study we suppose that the smooth vector fields $\eta, \eta_1,\eta_2\cdots,$ $ \iota, \iota_1,\iota_2\cdots \in \chi(M)$.  Now, corresponding to an endomorphism $\mathcal{I}(\iota_1,\iota_2),$ a $(0,4)$-tensor $I$ can be defined  as
$$I(\iota_1,\iota_2,\iota_3,\iota_4)=g(\mathcal{I}(\iota_1,\iota_2)\iota_3,\iota_4).$$
If the corresponding endomorphism $\mathcal{I}$ is replaced by $\mathcal{I}_{R}$ (resp., $\mathcal{I}_{C}$, $\mathcal{I}_{K}$, $\mathcal{I}_{W}$ and $\mathcal{I}_{P}$), the $(0,4)$-tensor $I$ turns into the  Riemann curvature tensor  $R$  (resp.,  conformal curvature $C$, conharmonic curvature $K$, concircular curvature $W$ and  projective curvature $P$).\\
\indent On a  $(0,r)$-tensor $\zeta$, $r\ge 1$, we simulate an endomorphism $\mathcal{I}(\iota_1,\iota_2)$ to define   $(0,r+2)$-type tensor  $I\cdot \zeta$ given as follows (\cite{DG02,DGHS98,DH03,SDHJK15,SK16}):
\beb
(I \cdot \zeta) (\eta_1,\eta_2,\cdots,\eta_r,\iota_1,\iota_2) &=&(\mathcal{I}(\iota_1,\iota_2) \zeta)(\eta_1,\eta_2,\cdots,\eta_r)\\ &=& - \zeta(\mathcal{I}(\iota_1,\iota_2)\eta_1,\eta_2,\cdots, \eta_r) - \cdots - \zeta(\eta_1,\eta_2,\cdots,\mathcal{I}(\iota_1,\iota_2)\eta_r).
	\eeb
If we take $\mathcal{I}(\iota_1,\iota_2)=\iota_1\wedge_{\nu} \iota_2,$ then the  $(0,r+2)$-type tensor $Q(\nu, \zeta)$ is  known as Tachibana tensor defined as (\cite{DGPSS11,SDHJK15,SK14,Tach74})
\beb
&&Q(\nu,\zeta)(\eta_1,\eta_2,\cdots , \eta_r,\iota_1,\iota_2) = ((\iota_1 \wedge_{\nu} \iota_2)  \zeta)(\eta_1,\eta_2, \cdots ,\eta_r)\\
&&\hspace{1.5in} = \nu(\iota_1,\iota_1)\zeta(\iota_2,\eta_2,\cdots, \eta_r) + \cdots + \nu(\iota_1,\iota_r)\zeta(\eta_1,\eta_2, \cdots , \iota_2)\\
&&\hspace{1.5in} - \nu(\iota_2,\eta_1)\zeta(\iota_1,\eta_2, \cdots, \eta_r) - \cdots - \nu(\iota_2,\eta_r)\zeta(\eta_1,\eta_2, \cdots, \iota_1).
\eeb

In terms of the local coordinates, the tensor $I\cdot \zeta$ and the Tachibana tensor $Q(\nu, \zeta)$ can be rewritten as
\beb
(I\cdot \zeta)_{b_1b_2...b_r\alpha \beta} &=& -g^{uv}[I_{\alpha \beta b_1v}\zeta_{ub_2...b_r} +\cdots + I_{\alpha \beta b_lv}\zeta_{b_1b_2...u}],\\
Q(\nu,\zeta)_{b_1b_2...b_r\alpha \beta} &=& \nu_{b_1 \beta }\zeta_{\alpha b_2...b_r} + \cdots + \nu_{b_r \beta }\zeta_{b_1b_2...\alpha} \\ 
&-& \nu_{b_1 \alpha}\zeta_{\beta b_2...b_r} - \cdots - \nu_{b_r \alpha}\zeta_{b_1b_2...\beta}.
\eeb
%========================================================================
\begin{defi} \cite{AD83,Desz92,Desz93,DGHS00,DGHZ15,DGHZ16,SK14,SK18,SKppsnw}	If the condition $I\cdot \zeta=f_{_\zeta}Q(g,\zeta)$ holds for a smooth scalar function $ f_{_\zeta} $ on $M$, i.e., the tensors $I \cdot \zeta$ and $Q(g, \zeta)$ are linearly dependent on  $M$, then $M$ is called a $\zeta$-pseudosymmetric manifold due to the tensor $I$. %($f_{_\zeta}$ being some smooth  function on $M$) so, this is known as $\zeta$- pseudosymmetric manifold due to  $I$.
	 Also, if the tensors $I\cdot \zeta$ and $Q(S, \zeta)$ are linearly dependent by the relation $I\cdot \zeta=\widetilde{f}_{_\zeta}Q(S,\zeta)$  with  a smooth scalar function $\widetilde{f}_{_\zeta}$ on  $M$, then $M$ is called a Ricci generalized $\zeta$-pseudosymmetric manifold due to the tensor $I$. In particular, a $\zeta$-semisymmetric manifold due to the tensor $I$ is defined by the relation $I \cdot \zeta=0$. 
\end{defi}
\indent In the relation $I\cdot \zeta=f_{_\zeta}Q(g,\zeta)$  if $I=\zeta=R$, %(resp., $S$, $C$, $K$, $W$ and $P$), 
then $M$ is simply called a pseudosymmetric  %(resp.,Ricci, conformally, conharmonically and concircularly and projective) 
manifold  and  for $I=R$ and $\zeta=K$ (resp., $S$,  $P$, $W$ and $C$),  it is called conharmonic (resp., Ricci, projective, concircular and conformal) pseudosymmetric manifold.  Similarly, various  types of  Ricci generalized pseudosymmetric  and  semisymmetric  manifolds  can be obtained by considering $I$ and $\zeta$ as others curvature tensors.\\

\indent Again, if the Ricci tensor $S$ is proportional to the metric tensor $g$ on $M$, i.e.,  $S=\frac{ \kappa}{n} g$, then $M$ is said to be an Einstein manifold \cite{Bess87}, and the manifold $M$ is called $m$-quasi-Einstein  \cite{S09, SKH11, SK19,SYH09} if the rank of $(S-\alpha g)$ is $m$ for some scalar $\alpha$, and in this case, Ricci tensor locally takes the form $S=\alpha g+ \beta \Gamma \otimes \Gamma$ with some scalars $\alpha,\beta$  and $1$-form $\Gamma$. Also, if $\alpha=0$, then the $m$-quasi-Einstein manifold turns into a Ricci simple manifold. We note that Morris-Thorne  spacetime \cite{ECS22} and G\"{o}del spacetime \cite{DHJKS14} are Ricci simple manifolds, Robertson-Walker spacetime \cite{ADEHM14} and Siklos spacetime \cite{SDKC19}  are quasi-Einstein manifolds, Kantowski-Sachs spacetime \cite{SC21} and Som-Raychaudhuri spacetime \cite{SK16} are $2$-quasi Einstein manifolds  and Kaigorodov spacetime \cite{SDKC19} is an Einstein manifold. For curvature properties of Robinson-Trautman metric, Melvin magnetic metric  and generalized pp-wave metric, etc., we refer the reader to see \cite{EC21,SAA18,SAAC20N,SDC}. 
 
%\begin{defi}$($\cite{S09,SK19}$)$ 
	
%	A semi-Riemannian manifold $M$
%	If on a manifold $M$, is called k-quasi Einstein manifold if rank of $(S-\alpha g)=k$, $\alpha$ being scalar, $0\leq k\leq (n-1)$. The  manifold is called quasi-Einstein (resp., $2$-quasi Einstein and Einstein) if $k=1$ (resp., $k=2$ and $k=1$).  If $\alpha=0$, than the   quasi-Einstein manifold turns into  Ricci simple manifold.  \\

%It is to be noted that Morris-Thorne  spacetime \cite{ECS22} is a Ricci simple manifold, Robertson-Walker spacetime \cite{ADEHM14} is quasi-Einstein, Kantowski-Sachs spacetime \cite{SC21} is $2$-quasi Einstein and Kaigorodov spacetime \cite{SDKC19} is Einstein.

%\end{defi}

\begin{defi}\label{def_GQE} \cite{C01}
	The manifold $M$ is said to be generalized quasi-Einstein if 
	$$S=\alpha g+\beta \Theta \otimes \Theta +\gamma (\Theta \otimes \Sigma+\Sigma \otimes \Theta)$$ holds 
	%on $U=\{\ell\in M : \big(S-\alpha g-\beta \Pi \otimes \Pi -\gamma (\Pi \otimes \phi+\phi \otimes \Pi)\big)_\ell \neq 0 \text{ with scalars  $\alpha,\beta,\gamma$ and 1-forms $\Pi,\phi$} \}$ 
	for some smooth scalar functions $\alpha$, $\beta$, $\gamma$ and mutually orthogonal $1$-forms $\Theta$ and $\Sigma$.
\end{defi}	
%It is to be noted that Som-Raychaudhury spacetime is a 
In the literature, there are other notions of generalized quasi-Einstein manifolds (see, Shaikh \cite{S09}). But throughout the paper we will consider the generalized quasi-Einstein manifold by Chaki \cite{C01} as given in Definition \ref{def_GQE}.

\begin{defi} $($\cite{Bess87,SK14,  SK16srs,SK19}$)$
	If $S^2, S^3, S^4$, defined by $S^{\lambda+1}(\iota_1,\iota_2)=S^\lambda(\iota_1,\mathscr E\iota_2)$ with $\lambda=1,2,3$, are linearly dependent by the relation %linear dependency
	$$\aleph_1g +\aleph_2 S + \aleph_3 S^2 + \aleph_4 S^3 + S^4  =0 $$ $$(resp.,\ \  \aleph_5 g + \aleph_6 S +\aleph_7 S^2 + S^3=0 \; \text{ and }\;   \aleph_8 g +\aleph_9 S + S^2=0)$$
	on $M$ for some scalar functions $\aleph_i$ ($1\le i\le 9$), then $M$ is called an Ein$(4)$ (resp., Ein$(3)$ and Ein$(2)$) manifold.
\end{defi}
It is noteworthy to mention that Melvin magnetic spacetime \cite{SAAC20} and Siklos spacetime \cite{SDKC19} are Ein$(2)$ manifolds, while Lifshitz spacetime \cite{SSC19}  and Som-Raychaudhuri spacetime \cite{SK16} are Ein$(3)$ manifolds.

\begin{defi} \cite{Desz03,DGJPZ13,DGJZ-2016, DGP-TV-2015, SDHJK15, SK16,SK19}
	 If the Riemann-Christoffel curvature tensor $R$ can be expressed as a linear combination of the tensors $g\wedge g$, $g\wedge S$, $S\wedge S$, $g\wedge S^2$, $S\wedge S^2$ and $S^2 \wedge S^2$, given by
	
	$$R=(\mathscr{B}_{1} S^2 + \mathscr{B}_{2}  S + \mathscr{B}_{3} g) \wedge S^2 + (\mathscr{B}_{4} S + \mathscr{B}_{5} g) \wedge S + \mathscr{B}_{6} (g \wedge g)$$
	
		$$(resp.,\ \  R=(\mathscr{B}_{7} S+\mathscr{B}_{8}g)\wedge S + \mathscr{B}_{9} g\wedge g)$$
	for some  scalars $\mathscr{B}_{i}$, $1\leq i \leq 9$, then $M$ is called a generalized Roter type (resp., Roter type \cite{Desz03, DG02,  DGP-TV-2011, DPSch-2013, Glog-2007}) manifold.

\end{defi}
We mention that Vaidya-Bonner spacetime \cite{SDC} and Lifshitz spacetime \cite{SSC19} are generalized Roter type manifold, and   Nariai spacetime \cite{SAAC20N} and Melvin magnetic spacetime \cite{SAAC20} are Roter type manifold.

%In \cite{DDH21} it is proven that the Roter type condition on $4$-dimensional Lorentzian manifolds is essentially equivalent with the condition of pseudosymmetry and Nariai spacetime \cite{SAAC20N} is a Roter type spacetime while Lifshitz spacetime \cite{SSC19} and Vaidya-Bonner spacetime \cite{SDC21} are examples of generalized Roter type manifolds.

\begin{defi} \label{def_WSym} \cite{TB89, TB93}
	A manifold $M$ is called a weakly symmetric in the sense of Tam\'{a}ssy and Binh  if the covariant derivative of Riemann curvature tensor $R$ can be expressed in the form
	\beb
	(\nabla_{X} R)(\eta_1,\eta_2,\eta_3,\eta_4)&=& \Pi(X)\otimes R(\eta_1,\eta_2,\eta_3,\eta_4)+ \Phi(\eta_4)\otimes R(\eta_1,\eta_2,\eta_3,X)\\ &+& \Phi(\eta_3)\otimes R(\eta_1,\eta_2,X,\eta_4)+ \Psi(\eta_2)\otimes R(\eta_1,X,\eta_3,\eta_4)\\ &+& \Psi(\eta_1)\otimes R(X,\eta_2,\eta_3,\eta_4), 
	\eeb
where $\Pi,\Phi$ and $\Psi$ are associated $1$-forms on $ M $. In particular, if $\Pi=2\Phi=2\Psi$, it is a Chaki pseudosymmetric manifold \cite{Chak87, Chak88}.
\end{defi}

\begin{defi}
The Ricci tensor of a manifold $M$ is  cyclic parallel (see, \cite{ Gray78, SB08, SS06, SS07}) if  $$(\nabla_{\eta_{1}}S)(\eta_2,\eta_3) + (\nabla_{\eta_{2}}S)(\eta_3,\eta_1) + (\nabla_{\eta_{3}}S)(\eta_1,\eta_2)=0$$  holds, and Codazzi type if the Ricci tensor  realizes the relation (see, \cite{F81, S81}) $$(\nabla_{\eta_{1}}S)(\eta_2,\eta_3) = (\nabla_{\eta_{2}}S)(\eta_1,\eta_3).$$
	\end{defi}
We note that the Ricci tensor of G\"odel spacetime \cite{DHJKS14} is  cyclic parallel and the Ricci tensor of $(t-z)$-type plane wave metric \cite{EC21} is of Codazzi type.
\begin{defi} $($\cite{DD91,DGHS98,DGJPZ13, DGPSS11,MM12a, MM12b, MM14}$)$
	
	Let $\zeta$ be a  $(0,4)$-type tensor on $M$. Then a symmetric (0,2)-type tensor $\nu$ corresponding to the endomorphism $\mathcal I_\nu$ is said to be  $\zeta$-compatible  if
	\[
	\zeta(\mathcal I_\nu \eta_1, \iota,\eta_2,\eta_3) + \zeta(\mathcal I_\nu \eta_2, \iota,\eta_3,\eta_1) + \zeta(\mathcal I_\nu \eta_3, \iota,\eta_1,\eta_2) = 0,
	\]
	holds on $M$. Again, if $\varphi \otimes \varphi$ is $\zeta$-compatible for an 1-form $\varphi$, then $\varphi$  is called a $\zeta$-compatible form. 
\end{defi}
Replacing $\zeta$ by the curvature tensor $R$ (resp., $C$, $W$, $P$ and $K$), the Riemann (resp., conformal, concircular, projective and conharmonic)  compatibility of $\nu$ can be obtained.

\begin{defi} 
	For a  tensor $I$ of type $(0,4)$, the curvature $2$-forms $\Omega^m_{(I)}l$ \cite{SKP03} are called recurrent \cite{ LR89, MS12a, MS13a, MS14} if 
	\[\mathop{\mathcal{S}}_{\eta_1,\eta_2,\eta_3}(\nabla_{\eta_{1}} I) (\eta_2,\eta_3,\iota,\eta) = \mathop{\mathcal{S}}_{\eta_1,\eta_2,\eta_3}\sigma (\eta_1) I(\eta_2,\eta_3,\iota,\eta)\]
	holds on $M$,
	where $\mathcal{S}$ is the cyclic sum over $\eta_1, \eta_2, \eta_3$    and for a $(0,2)$ tensor field $\nu$, the $1$-forms $\wedge_{(\nu)l}$ (\cite{SKP03}) are called recurrent if 
	$(\nabla_{\eta_1} \nu) (\eta_2,\iota) - (\nabla_{\eta_2} \nu) (\eta_1,\iota) = \sigma (\eta_1) \nu(\eta_2,\iota) - \sigma (\eta_2) \nu(\eta_1,\iota) $, for some $1$-form $\sigma$.
	
\end{defi}

\begin{defi} %$($\cite{P95,SK16,Venz85}$)$ 
	$($\cite{ P95, SK16, Venz85}$)$
	For a $(0,4)$-type tensor $I$ if $M$ admits the relation 
	\[\mathop{\mathcal{S}}_{\eta_1,\eta_2,\eta_3}\sigma (\eta_1)\otimes I(\eta_2,\eta_3,\iota,\eta) = 0,\]
	where $\mathcal{S}$ is the  cyclic sum over $\eta_1, \eta_2, \eta_3$   and $\mathcal{L}(M)$ is the vector space of all $1$-forms with dimension $\geq 1$, then $M$ is called $I$-space by Venzi.
\end{defi}

Now, we give some definitions of geometrical symmetries, such as, motion, curvature collineation, curvature inheritance, Ricci collineation and Ricci inheritance, which are originated from the Lie derivatives of several tensors.  

\begin{defi}
	A manifold $M$ admits motion with respect to some vector field $\eta$ if $\pounds_\eta g=0$, where $\pounds_\eta$ represents the Lie derivative with respect to $\eta$.
\end{defi}

 In 1969, Katzin \textit{et al.} \cite{KLD1969,KLD1970} defined the notion of curvature collineation by vanishing Lie derivative of the Riemann curvature tensor with respect to some vector field. Again, in 1992, Duggal \cite{Duggal1992} generalizes the concept of curvature collineation by introducing the notion of curvature inheritance.
 
 \begin{defi}\label{def_CI} (\cite{Duggal1992})
 	A manifold $M$ possesses curvature inheritance if there is a vector field $\eta$ which satisfies
 	$$\pounds_\eta \widetilde R=\lambda \widetilde R,$$
 	where $\lambda$ is a scalar function and the (1,3)-type curvature tensor $\widetilde R$ is associated with the (0,4)-type curvature tensor $R$  by the relation  $R(\iota_1,\iota_2,\iota_3,\iota_4)=g(\widetilde R(\iota_1,\iota_2)\iota_3,\iota_4)$. In particular, if $\lambda=0$, i.e., $\pounds_\eta \widetilde R=0$, then it turns into curvature collineation \cite{KLD1969,KLD1970}.
 \end{defi}

 \begin{defi} \label{def_RI}(\cite{Duggal1992})
	A manifold $M$ admits Ricci inheritance if it realizes the relation
		$$\pounds_\eta S=\lambda S$$
	for some vector field $\eta$ and scalar function $\lambda$. Further, if $\lambda=0$, it turns into Ricci collineation (i.e., $\pounds_\eta S=0$).
\end{defi}

Again, recently Shaikh and Datta \cite{ShaikhDatta2022} introduced the notion of generalized curvature inheritance, which is defined as follows:
\begin{defi}\label{def_GCI} (\cite{ShaikhDatta2022})
		A manifold $M$ admits generalized curvature inheritance if there is a vector field $\eta$ which possesses
		 	$$\pounds_\eta  R=\lambda  R + \lambda_1 g\wedge g +\lambda_2 g\wedge S +\lambda_3 S\wedge S ,$$
		 where $\lambda,\lambda_1,\lambda_2,\lambda_3$ are the scalar functions. In particular, if $\lambda_i=0$ for $i=1,2,3$, then it $M$ admits curvature inheritance. Further, if  $\lambda=0=\lambda_i$ for $i=1,2,3$, then it turns into curvature collineation.
\end{defi}
At first glance, it seems that the notion of curvature inheritance in  Definition \ref{def_GCI} resembles the Definition \ref{def_CI}. But, they are completely different ideas indeed, as $\widetilde R\neq R$. In this paper, it is also proved  that the notions of curvature inheritance in Definition \ref{def_CI}  and Definition \ref{def_GCI} are not equivalent as shown by PGM spacetime.

%%%%%%%%%%%%%%%%%%%%%%%%%%%%%%%%%%%%%%%%%%%%%%%%%%%%%%%%%%%%%%%%%%%%%%%%%%%%%%%%%%%%
\section{\bf Point-like global monopole spacetime admitting curvature related geometric structures}\label{com}
%%%%%%%%%%%%%%%%%%%%%%%%%%%%%%%%%%%%%%%%%%%%%%%%%%%%%%%%%%%%%%%%%%%%%%%%%%%%%%%%%%%%

The line-element of PGM spacetime which is a static and spherically symmetric metric in the spherical coordinates $(t=x^1, r=x^2, \theta=x^3, \phi=x^4)$ is given by ($c=\hbar=G$) \cite{MBAV,ERFM2,CF44,RV,GAM,PN,FA,FA2,FA3,FA4,FA5,FA6,FA7,FA8,FA9,FA10}
\bea\label{GMS}
ds^2=-dt^2+\frac{dr^2}{\alpha^2}+r^2\,(d\theta^2+\sin^2 \theta\,d\phi^2)=g_{\mu\nu}\,dx^{\mu}\,dx^{\nu},
\eea
where the nonzero components of the covariant and contravariant metric tensors are $g_{11}=-1=g^{11}$,  $g_{22}=\frac{1}{\alpha^2}$, $g_{33}=r^2$, $g_{44}=r^2 \sin^2 \theta $ and $g_{\mu\nu}=0$, $\mu \neq \nu$ for $\mu,\nu=1,2,3,4$.

The non-vanishing components $\Gamma^{\gamma}_{\mu\nu}$ of the Christoffel symbols of the second kind are given by
$$\label{chis}\begin{array}{c}
\Gamma^3_{23}=\frac{1}{r}=\Gamma^4_{24},\quad \Gamma^2_{33} =-r \alpha^2,\quad \Gamma^4_{34}=\cot\theta, \quad \Gamma^2_{44}=-r \alpha^2 \sin^2 \theta,\quad \Gamma^3_{44}=-\cos\theta \sin\theta 
 .\\
\end{array}$$

%--------------
The non-vanishing components (upto symmetry) of Riemann curvature tensor  $R_{\mu\nu\sigma\lambda}$ and  the Ricci tensor of $S_{\mu\nu}$ are obtained as follows:
%---------------
\begin{equation}
\label{R04}\begin{array}{c}
R_{3434}=-r^2 (-1+\alpha^2) \sin^2  \theta ; 
\end{array}
\end{equation}
\begin{equation}\label{S02}
	\begin{array}{c}
S_{33}=-1+\alpha^2,  \
S_{44}=(-1+\alpha^2) \sin^2 \theta. \\
\end{array}
\end{equation}

The scalar curvature $\kappa$ is given by   $\kappa=\frac{2 (-1+\alpha^2)}{r^2}$.\\
This leads to the following:
\begin{pr} \label{pr31}
	The PGM spacetime \eqref{GMS} is neither Einstein  nor quasi-Einstein manifold but
	\begin{enumerate}[label=(\roman*)] 
		%----------------------------------------------------------
		\item it is an  Einstein manifold of degree  $2$, i.e., it fulfills the condition $S^2 = \frac{(-1+\alpha^2)}{r^2}S$,
		\item it is $2$-quasi Einstein and generalized quasi-Einstein manifold,
		\item its Riemann curvature can be decomposed by $R= \frac{r^2}{2(-1+\alpha^2)} S\wedge S$,
		%-----------------------------------------------------------
		\item its Ricci tensor  is Riemann compatible, conharmonic compatible, concircular compatible,  projective compatible and conformal compatible.

	\end{enumerate}

\end{pr}

%------------------
Let $\mathcal{V}^1= \nabla R $ and  $\mathcal{V}^2=\nabla S $.
Then the non-vanishing components (upto symmetry) of the covariant derivatives of the Riemann curvature tensor $R$ and the Ricci tensor  $S$ are given by 
$$\begin{array}{c}
\mathcal{V}^1_{2334,4}=-r (-1+\alpha^2) \sin^2 \theta =-\mathcal{V}^1 _{2434,3}, \ 
\mathcal{V}^1_{3434,2}=2 r (-1+\alpha^2) \sin^2 \theta, \\ 

\mathcal{V}^2_{23,3}=\frac{1-\alpha^2}{r}, \mathcal{V}^2_{24,4}=-\frac{(-1+\alpha^2) \sin^2 \theta}{r}, \
\mathcal{V}^2_{33,2}=\frac{2 -2\alpha^2}{r}, \mathcal{V}^2_{44,2}=-\frac{2 (-1+\alpha^2) \sin^2 \theta}{r}. \\

\end{array}$$
\indent  The  components (upto symmetry) other than zero  of the  conformal curvature tensor $C$  are given below:
$$\begin{array}{c}
C_{1212}=\frac{-1+\alpha^2}{3 r^2 \alpha^2}, C_{1313}=\frac{1-\alpha^2}{6}, C_{1414}=-\frac{(-1+\alpha^2) \sin^2 \theta}{6}, \\
C_{2323}=\frac{1}{6}-\frac{1}{6\alpha^2}, C_{2424}=\frac{(-1+\alpha^2)\sin^2 \theta}{6\alpha^2}, 
C_{3434}=-\frac{r^2 (-1+\alpha^2) \sin^2 \theta }{3}. \\

\end{array}$$
\indent  The  components (upto symmetry) other than zero  of the  projective curvature tensor $P$  are shown as follows:
$$\begin{array}{c}
P_{1331}=\frac{-1+\alpha^2}{3 },  P_{1441}=\frac{1}{3} (-1+\alpha^2) \sin^2 \theta, P_{2442}=-\frac{(-1+\alpha^2)\sin^2\theta}{3\alpha^2},\\ P_{2332}=\frac{1}{3}(-1+\frac{1}{\alpha^2}),
P_{3434}=-\frac{2}{3} r^2 (-1+\alpha^2) \sin^2\theta=-P_{3443}. \\

\end{array}$$

\indent If $\mathcal{V}^3= \nabla C $, then the  components other than zero  of the  covariant derivative of conformal curvature tensor $C$ are given by 
$$\begin{array}{c}
\mathcal{V}^3_{1212,2}=-\frac{2(-1+\alpha^2)}{3r^3\alpha^2}, \mathcal{V}^3_{1213,3}=\frac{-1+\alpha^2}{2r}, 
\mathcal{V}^3_{1214,4}=\frac{(-1+\alpha^2) \sin^2 \theta}{2r}, \\ 
\mathcal{V}^3_{1313,2}=\frac{-1+\alpha^2}{3r}, 
\mathcal{V}^3_{1414,2}=\frac{(-1+\alpha^2) \sin^2 \theta}{3r},
\mathcal{V}^3_{2323,2}=-\frac{-1+\alpha^2}{3r\alpha^2},  \mathcal{V}^3_{2424,2}=-\frac{(-1+\alpha^2)\sin^2 \theta}{3r\alpha^2}, \\
\mathcal{V}^3_{2334,4}=-\frac{r (-1+\alpha^2) \sin^2 \theta}{2}=-\mathcal{V}^3_{2434,3}, 

\mathcal{V}^3_{3434,2}=\frac{2 r (-1+\alpha^2) \sin^2 \theta}{3} . 
\end{array}$$

From the above tensor components, we can state the following:
\begin{pr} \label{pr32}
	 The   PGM spacetime \eqref{GMS}  realizes the following:
	\begin{enumerate}[label=(\roman*)] 
			\item  its Ricci $1$-forms are recurrent, i.e., $\nabla_{\eta_1} S(\eta_2,\eta_3)- \nabla_{\eta_2} S(\eta_1,\eta_3)= \vartheta(\eta_1)\otimes S(\eta_2,\eta_3)-\vartheta(\eta_2)\otimes S(\eta_1,\eta_3)$ for $\vartheta=\left\lbrace 0,-\frac{1}{r},0,0\right\rbrace $,
			%-----------------------------------------------------------
			\item its conformal curvature $C$ is recurrent for the 1-form $\left\lbrace0,\frac{1}{r},0,0 \right\rbrace $,
			%-----------------------------------------------------------
			\item it is a $R$-space by Venzi  for $\left\lbrace0,0,1,1 \right\rbrace $, 
			\item it is Chaki pseudosymmetric for the  1-form $\Pi=\left\lbrace0,-\frac{1}{r},0,0 \right\rbrace $,
			
			\item it is semisymmetric as $R\cdot R=0$. Therefore, it is Ricci semisymmetric, conharmonic semisymmetric, projective semisymmetric, concircular semisymmetric and  conformal semisymmetric, and hence it is also pseudosymmetric, Ricci pseudosymmetric, conformal pseudosymmetic in the sense of Deszcz.
	\end{enumerate}

\end{pr}

%\begin{cor}
%	The point-like global monopole metric also admits weakly symmetric for the $1$-form $\Pi=\left\lbrace0,-\frac{2}{r},0,0 \right\rbrace $ of Riemann and Ricci. 
%\end{cor}

Let $\mathcal{Z}^1= C\cdot R$, $\mathcal{Z}^2= C\cdot C$, $\mathcal{Z}^3 = P\cdot C$, $\mathcal{H}^1= Q(g,R)$, $\mathcal{H}^2= Q(g,C)$ and $\mathcal{H}^3= Q(S,C)$. Then the components other than zero of $\mathcal{Z}^1$, $\mathcal{Z}^2$, $\mathcal{Z}^3$, $\mathcal{H}^1$, $\mathcal{H}^2$ and $\mathcal{H}^3$ are computed as follows:
$$\begin{array}{c}
\mathcal{Z}^1_{1434,13}= -\frac{(-1+\alpha^2)^2 \sin^2 \theta}{6}= -\mathcal{Z}^1_{1334,14},
\mathcal{Z}^1_{2434,23}= \frac{(-1+\alpha^2)^2 \sin^2 \theta}{6\alpha^2}= -\mathcal{Z}^1_{2334,24}; 
\end{array}$$
$$\begin{array}{c}
\mathcal{Z}^2_{1223,13}= -\frac{(-1+\alpha^2)^2 }{12r^2 \alpha^2}= -\mathcal{Z}^2_{1213,23}, 
\mathcal{Z}^2_{1434,13}= -\frac{(-1+\alpha^2)^2 \sin^2 \theta}{12}= -\mathcal{Z}^2_{1334,14}, \\
\mathcal{Z}^2_{1224,14}=-\frac{(-1+\alpha^2)^2 \sin^2 \theta}{12r^2 \alpha^2}= -\mathcal{Z}^2_{1214,24}, 
\mathcal{Z}^2_{2434,23}= \frac{(-1+\alpha^2)^2 \sin^2 \theta}{12\alpha^2}= -\mathcal{Z}^2_{2334,24}; 
\end{array}$$
$$\begin{array}{c}
\mathcal{Z}^3_{1223,13}= -\frac{(-1+\alpha^2)^2 }{9r^2\alpha^2}= -\mathcal{Z}^3_{1213,23}, 
\mathcal{Z}^3_{1434,13}= -\frac{(-1+\alpha^2)^2 \sin^2 \theta}{18}= -\mathcal{Z}^3_{1334,14}, \\
\mathcal{Z}^3_{1224,14}=-\frac{(-1+\alpha^2)^2 \sin^2 \theta}{9r^2\alpha^2}=- \mathcal{Z}^3_{1214,24}, 
\mathcal{Z}^3_{2434,23}= \frac{(-1+\alpha^2)^2 \sin^2 \theta}{18\alpha^2}= -\mathcal{Z}^3_{2334,24}, \\
\mathcal{Z}^3_{1223,31}= \frac{(-1+\alpha^2)^2 }{9r^2\alpha^2}= -\mathcal{Z}^3_{1213,32}, 
\mathcal{Z}^3_{1434,31}= \frac{(-1+\alpha^2)^2 \sin^2 \theta}{18}= -\alpha^2\mathcal{Z}^3_{2434,32}, \\
\mathcal{Z}^3_{1224,41}=\frac{(-1+\alpha^2)^2 \sin^2 \theta}{9r^2\alpha^2}= -\mathcal{Z}^3_{1214,42}, 
\mathcal{Z}^3_{1334,41}= -\frac{(-1+\alpha^2)^2 \sin^2 \theta}{18}=-\alpha^2\mathcal{Z}^3_{2334,42}; 
\end{array}$$
$$\begin{array}{c}
\mathcal{H}^1_{1434,13}=r^2 (-1+\alpha^2) \sin^2 \theta=-\mathcal{H}^1_{1334,14}, 
\mathcal{H}^1_{2434,23}=-\frac{r^2(-1+\alpha^2)\sin^2\theta}{\alpha^2} =-\mathcal{H}^1_{2334,24}; 
\end{array}$$
$$\begin{array}{c}
\mathcal{H}^2_{1223,13}= \frac{1 }{2}-\frac{1}{2\alpha^2}= -\mathcal{H}^2_{1213,23}, 
\mathcal{H}^2_{1434,13}= \frac{ r^2 (-1+\alpha^2) \sin^2 \theta}{2}= -\mathcal{H}^2_{1334,14}, \\
\mathcal{H}^2_{1224,14}=\frac{(-1+\alpha^2) \sin^2 \theta}{2\alpha^2}= -\mathcal{H}^2_{1214,24}, 
\mathcal{H}^2_{2434,23}= -\frac{r^2(-1+\alpha^2) \sin^2 \theta}{2\alpha^2}= -\mathcal{H}^2_{2334,24}; 
\end{array}$$
$$\begin{array}{c}
\mathcal{H}^3_{1223,13}= \frac{(-1+\alpha^2)^2 }{3r^2\alpha^2}= -\mathcal{H}^3_{1213,23}, 
\mathcal{H}^3_{1434,13}= \frac{(-1+\alpha^2)^2 \sin^2 \theta}{6}= -\mathcal{H}^3_{1334,14}, \\
\mathcal{H}^3_{1224,14}=\frac{(-1+\alpha^2)^2 \sin^2 \theta}{3r^2\alpha^2}= -\mathcal{H}^3_{1214,24}, 
\mathcal{H}^3_{2434,23}= -\frac{(-1+\alpha^2)^2 \sin^2 \theta}{6\alpha^2}= -\mathcal{H}^3_{2334,24}.
\end{array}$$
The above calculation of tensors leads to the following:

\begin{pr}\label{pr33}
	The PGM spacetime \eqref{GMS}  satifies the pseudosymmetric type curvature conditions
	$$C\cdot R=-\frac{(-1+\alpha^2)}{6r^2}Q(g,R),\ \ \   C\cdot C=-\frac{(-1+\alpha^2)}{6r^2}Q(g,C)\  \text{ and }\ P\cdot C=-\frac{1}{3}Q(S,C),$$
i.e., it is pseudosymmetric due to conformal curvature tensor, pseudosymmetric Weyl curvature tensor and also Ricci generalized conformal peudosymmetric due to projective curvature tensor.
\end{pr}

\indent The  components other than zero  of the  concircular curvature tensor $W$ of PGM spacetime  are given by
$$\begin{array}{c}
W_{1212}=-\frac{(-1+\alpha^2)}{6 r^2 \alpha^2}, W_{1313}=-\frac{(-1+\alpha^2)}{6}, W_{1414}=-\frac{(-1+\alpha^2) \sin^2 \theta}{6}, \\
W_{2323}=\frac{1}{6}-\frac{1}{6\alpha^2}, W_{2424}=\frac{(-1+\alpha^2)\sin^2 \theta}{6\alpha^2}, 
W_{3434}=-\frac{5 r^2 (-1+\alpha^2) \sin^2 \theta }{6}. \\

\end{array}$$

\indent If $\mathcal{V}^4= \nabla W $, then the   components other than zero  of the  covariant derivative of concircular curvature tensor $W$ are given by 
$$\begin{array}{c}
\mathcal{V}^4_{1212,2}=\frac{-1+\alpha^2}{3r^3 \alpha^2},\mathcal{V}^4_{1313,2}=\frac{-1
	+\alpha^2}{3r}, 
\mathcal{V}^4_{1414,2}=\frac{(-1+\alpha^2) \sin^2 \theta}{3r}, \\ 
\mathcal{V}^4_{2323,2}=-\frac{(-1+\alpha^2)}{3r\alpha^2}, 
\mathcal{V}^4_{2334,4}=- r (-1+\alpha^2) \sin^2 \theta=-
\mathcal{V}^4_{2434,3},
\mathcal{V}^4_{2424,2}=-\frac{(-1+\alpha^2)\sin^2 \theta}{3r\alpha^2}, \\
\mathcal{V}^4_{3434,2}=\frac{5 (-1+\alpha^2) r  \sin^2 \theta}{3}.
\end{array}$$

Let $\mathcal{Z}^4= W\cdot R$, $\mathcal{Z}^5= P\cdot W$, $\mathcal{H}^4= Q(S,W)$. Then the components other than zero of the tensor $\mathcal{Z}^4$, $\mathcal{Z}^5$,  $\mathcal{H}^4$ are given as follows:

$$\begin{array}{c}
\mathcal{Z}^4_{1434,13}= -\frac{(-1+\alpha^2)^2 \sin^2 \theta}{6}= -\mathcal{Z}^4_{1334,14},
\mathcal{Z}^4_{2434,23}= \frac{(-1+\alpha^2)^2 \sin^2 \theta}{6\alpha^2}=- \mathcal{Z}^4_{2334,24};
\end{array}$$
$$\begin{array}{c}
\mathcal{Z}^5_{1223,13}= \frac{(-1+\alpha^2)^2 }{18r^2 \alpha^2}= \mathcal{Z}^5_{1213,32}, 
\mathcal{Z}^5_{1434,13}= -\frac{(-1+\alpha^2)^2 \sin^2 \theta}{18}= \mathcal{Z}^5_{1334,41}, \\
\mathcal{Z}^5_{1224,14}=\frac{(-1+\alpha^2)^2 \sin^2 \theta}{18r^2\alpha^2}= -\mathcal{Z}^5_{1214,42},
\mathcal{Z}^5_{1334,14}=\frac{(-1+\alpha^2)^2 \sin^2 \theta}{18}=\mathcal{Z}^5_{1434,31}, \\
\mathcal{Z}^5_{1213,23}=- \frac{(-1+\alpha^2)^2 }{18r^2\alpha^2}= \mathcal{Z}^5_{1223,31},
\mathcal{Z}^5_{2434,23}= \frac{(-1+\alpha^2)^2 \sin^2 \theta}{18\alpha^2}=- \mathcal{Z}^5_{2334,42},  \\
\mathcal{Z}^5_{1214,24}=-\frac{(-1+\alpha^2)^2 \sin^2 \theta}{18r^2\alpha^2}= \mathcal{Z}^5_{1224,41},
\mathcal{Z}^5_{2334,23}=- \frac{(-1+\alpha^2)^2 \sin^2 \theta}{18\alpha^2}= \mathcal{Z}^5_{2434,32},  \\
\end{array}$$

$$\begin{array}{c}
\mathcal{H}^4_{1223,13}=- \frac{(-1+\alpha^2)^2 }{6r^2\alpha^2}= -\mathcal{H}^4_{1213,23}, 
\mathcal{H}^4_{1434,13}= \frac{(-1+\alpha^2)^2 \sin^2 \theta}{6}= -\mathcal{H}^4_{1334,14}, \\
\mathcal{H}^4_{1224,14}=-\frac{(-1+\alpha^2)^2 \sin^2 \theta}{6r^2\alpha^2}= -\mathcal{H}^4_{1214,24}, 
\mathcal{H}^4_{2434,23}= -\frac{(-1+\alpha^2)^2 \sin^2 \theta}{6\alpha^2}= -\mathcal{H}^4_{2334,24}.
\end{array}$$
From the above calculation of tensors we can infer the following:
\begin{pr}\label{pr34}
	The PGM spacetime fulfills the curvature conditions
	$$\ W\cdot R=-\frac{(-1+\alpha^2)}{6r^2}Q(g,R) \ \ \text{ and }\ \ P\cdot W=-\frac{1}{3}Q(S,W)\ $$
	   i.e., the spacetime is pseudosymmetric due to concircular curvature tensor and also Ricci generalized concircular peudosymmetric due to projective curvature tensor.
\end{pr}

\indent The  components other than zero  of the  conharmonic curvature tensor $K$ of PGM spacetime  are given below:
$$\begin{array}{c}
K_{1313}=\frac{1-\alpha^2}{2}, K_{1414}=-\frac{(-1+\alpha^2) \sin^2 \theta}{2}, \\
K_{2323}=\frac{1}{2}-\frac{1}{2\alpha^2}, K_{2424}=\frac{(-1+\alpha^2)\sin^2 \theta}{2\alpha^2}. \\ 
\end{array}$$

\indent If $\mathcal{V}^5= \nabla K $, then the  components other than zero  of the  covariant derivative of conharmonic curvature tensor $K$ are given by
$$\begin{array}{c}
\mathcal{V}^5_{1213,3}=\frac{-1+\alpha^2}{2r}, \mathcal{V}^5_{1214,4}=\frac{(-1+\alpha^2) \sin^2 \theta }{2r}=\frac{1}{2}\mathcal{V}^5_{1414,2}, 
\mathcal{V}^5_{1313,2}=\frac{-1+\alpha^2 }{r}, \\ 
\mathcal{V}^5_{2323,2}=-\frac{1+\frac{1}{\alpha^2}}{r}, 
\mathcal{V}^5_{2334,4}=-  \frac{1}{2} r (-1+\alpha^2) \sin^2 \theta=-
\mathcal{V}^5_{2434,3},
\mathcal{V}^5_{2424,2}=-\frac{(-1+\alpha^2)\sin^2 \theta}{r\alpha^2}. \\
\end{array}$$

Let $\mathcal{Z}^6= K\cdot R$, $\mathcal{Z}^7= P\cdot K$, $\mathcal{Z}^8= P\cdot P$, $\mathcal{H}^5= Q(S,K)$ and $\mathcal{H}^6= Q(S,P)$. Then the components other than zero of the tensors $\mathcal{Z}^6$, $\mathcal{Z}^7$, $\mathcal{Z}^8$,  $\mathcal{H}^5$ and  $\mathcal{H}^6$ are computed as follows:
$$\begin{array}{c}
\mathcal{Z}^6_{1434,13}= -\frac{(-1+\alpha^2)^2 \sin^2 \theta}{2}= -\mathcal{Z}^6_{1334,14},
\mathcal{Z}^6_{2434,23}= \frac{(-1+\alpha^2)^2 \sin^2 \theta}{2\alpha^2}=- \mathcal{Z}^6_{2334,24}; 
\end{array}$$
$$\begin{array}{c}
\mathcal{Z}^7_{1434,13}= -\frac{(-1+\alpha^2)^2 \sin^2 \theta}{6}= -\mathcal{Z}^7_{1334,14}, 
\mathcal{Z}^7_{2434,23}=\frac{(-1+\alpha^2)^2 \sin^2 \theta}{6\alpha^2}=-\mathcal{Z}^7_{2334,24}, \\
\mathcal{Z}^7_{1434,31}= \frac{(-1+\alpha^2)^2 \sin^2 \theta}{6}=- \mathcal{Z}^7_{1334,41}, 
\mathcal{Z}^7_{2434,32}=-\frac{(-1+\alpha^2)^2 \sin^2 \theta}{6\alpha^2}=- \mathcal{Z}^7_{2334,42};
\end{array}$$
$$\begin{array}{c}
\mathcal{Z}^8_{1333,13}= \frac{(-1+\alpha^2)^2 }{9}= -\mathcal{Z}^8_{1333,31}, 
\mathcal{Z}^8_{1443,13}= \frac{(-1+\alpha^2)^2 \sin^2 \theta}{9}= \mathcal{Z}^8_{3441,13}=\mathcal{Z}^8_{1334,14}=\mathcal{Z}^8_{3431,41}, \\
\mathcal{Z}^8_{1444,14}=\frac{(-1+\alpha^2)^2 \sin^4 \theta}{9}= -\mathcal{Z}^8_{1444,41}, 
\mathcal{Z}^8_{3431,14}= -\frac{(-1+\alpha^2)^2 \sin^2 \theta}{9}= \mathcal{Z}^8_{1443,31}=\mathcal{Z}^8_{3441,31}=\mathcal{Z}^8_{1334,41}, \\
\mathcal{Z}^8_{2333,23}= -\frac{(-1+\alpha^2)^2 }{9\alpha^2}= -\mathcal{Z}^8_{2333,32}, 
\mathcal{Z}^8_{2443,23}= -\frac{(-1+\alpha^2)^2 \sin^2 \theta}{9\alpha^2}= \mathcal{Z}^8_{3442,23}=\mathcal{Z}^8_{2334,24}=\mathcal{Z}^8_{3432,24}, \\
\mathcal{Z}^8_{2444,24}= -\frac{(-1+\alpha^2)^2 \sin^4 \theta}{9\alpha^2}=-\mathcal{Z}^8_{2444,42}, 
\mathcal{Z}^8_{3432,24}= \frac{(-1+\alpha^2)^2 \sin^2 \theta}{9\alpha^2}= \mathcal{Z}^8_{2443,32}=\mathcal{Z}^8_{3442,32}=\mathcal{Z}^8_{2334,42};
\end{array}$$
$$\begin{array}{c}
\mathcal{H}^5_{1434,13}= \frac{(-1+\alpha^2)^2 \sin^2 \theta}{2}= -\mathcal{H}^5_{1334,14}, 
\mathcal{H}^5_{2434,23}= -\frac{(-1+\alpha^2)^2 \sin^2 \theta}{2\alpha^2}= -\mathcal{H}^5_{2334,24}; \\
\end{array}$$
$$\begin{array}{c}
\mathcal{H}^6_{1333,13}=-\frac{(-1+\alpha^2)^2}{3},
\mathcal{H}^6_{1443,13}=-\frac{(-1+\alpha^2)^2 \sin^2 \theta}{3} =\mathcal{H}^6_{3441,13}=\mathcal{H}^6_{1334,14}=-\mathcal{H}^6_{3431,14}, \\
\mathcal{H}^6_{1444,14}=-\frac{(-1+\alpha^2)^2 \sin^4 \theta}{3}, 
\mathcal{H}^6_{2333,23}=\frac{(-1+\alpha^2)^2}{3\alpha^2}, \\
\mathcal{H}^6_{2443,23}=\frac{(-1+\alpha^2)^2 \sin^2 \theta}{3\alpha^2} =\mathcal{H}^6_{3442,23}=\mathcal{H}^6_{2334,24}=-\mathcal{H}^6_{3432,24}, \\
\mathcal{H}^6_{2444,24}=\frac{(-1+\alpha^2)^2 \sin^4 \theta}{3\alpha^2};
\end{array}$$
The above computation of tensors leads to the following:
\begin{pr}\label{pr35}
	The PGM spacetime \eqref{GMS} fulfills the following pseudosymmetric type curvature conditions:
	$$\ K\cdot R=-\frac{(-1+\alpha^2)}{2r^2}Q(g,R), \ \ P\cdot K=-\frac{1}{3}Q(S,K)\ \text{ and } \ \ P\cdot P=-\frac{1}{3}Q(S,P), $$  
	i.e., the spacetime is pseudosymmetric due to conharmonic curvature tensor,  Ricci generalized conharmonic peudosymmetric due to projective curvature tensor and Ricci generalized projective pseudosymmetric.
\end{pr}

From  the above propositions, we can state that the PGM spacetime \eqref{GMS} admits  the following curvature restricted geometric properties:

\begin{thm}\label{mainthm}
The PGM spacetime \eqref{GMS} reveals the following curvature properties:
\begin{enumerate}[label=(\roman*)]
%\item it is Ricci simple as $S=\alpha \left( \eta\otimes \eta\right) $ holds for $\alpha=\frac{2b^2}{b^2+l^2}$, $\eta=\left\lbrace 0, 1, 0, 0 \right\rbrace $ with $\|\eta\|=1$, and hence it is special quasi-Einstein,
%------------------------------------------------------------------------ 
%\item it has special Ricci generalized pseudosymmetry as $R\cdot R = Q(S,R)$ is satisfied,
%----------------------------------------------------------------------------

\item it is pseudosymmetric due to conformal curvature tensor as  $C\cdot R = -\frac{(-1+\alpha^2)}{6r^2} Q(g,R)$. Hence 
    \begin{enumerate}[label=(\alph*)]
    \item $C\cdot S = -\frac{(-1+\alpha^2)}{6r^2} Q(g,S)$, 
    \item $C\cdot C = -\frac{(-1+\alpha^2)}{6r^2} Q(g,C)$ (i.e., pseudosymmetric Weyl conformal curvature tensor),
    \item $C\cdot W = -\frac{(-1+\alpha^2)}{6r^2} Q(g,W)$, 
    \item $C\cdot P = -\frac{(-1+\alpha^2)}{6r^2} Q(g,P)$ and 
    \item $C\cdot K = -\frac{(-1+\alpha^2)}{6r^2} Q(g,K)$,
    \end{enumerate}

\item it realizes pseudosymmetry due to concircular curvature tensor as  $W\cdot R = -\frac{(-1+\alpha^2)}{6r^2} Q(g,R)$. Hence 
    \begin{enumerate}[label=(\alph*)]
    \item $W\cdot S = -\frac{(-1+\alpha^2)}{6r^2} Q(g,S)$, 
    \item $W\cdot C = -\frac{(-1+\alpha^2)}{6r^2} Q(g,C)$, 
    \item $W\cdot W = -\frac{(-1+\alpha^2)}{6r^2} Q(g,W)$, 
    \item $W\cdot P = -\frac{(-1+\alpha^2)}{6r^2} Q(g,P)$ and 
    \item $W\cdot K = -\frac{(-1+\alpha^2)}{6r^2} Q(g,K)$,
    \end{enumerate}

\item it admits pseudosymmetry due to conharmonic curvature tensor as  $K\cdot R = -\frac{(-1+\alpha^2)}{2r^2} Q(g,R)$. Hence 
    \begin{enumerate}[label=(\alph*)]
    \item $K\cdot S = -\frac{(-1+\alpha^2)}{2r^2} Q(g,S)$, 
    \item $K\cdot C = -\frac{(-1+\alpha^2)}{2r^2} Q(g,C)$, 
    \item $K\cdot W = -\frac{(-1+\alpha^2)}{2r^2} Q(g,W)$, 
    \item $K\cdot P = -\frac{(-1+\alpha^2)}{2r^2} Q(g,P)$ and 
    \item $K\cdot K = -\frac{(-1+\alpha^2)}{2r^2} Q(g,K)$,
    \end{enumerate}

%------------------------------------------------------------------------------
\item it is Ricci generalized conformal pseudosymmetric due to projective curvature tensor as $P\cdot C=-\frac{1}{3}Q(S,C)$. Hence      
    \begin{enumerate}[label=(\alph*)]
    \item $P\cdot P=-\frac{1}{3}Q(S,P)$, 
    \item $P\cdot W=-\frac{1}{3}Q(S,W)$ and 
    \item $P\cdot K=-\frac{1}{3}Q(S,K)$,
    \end{enumerate}

%--------------------------------------------------------------
\item it is a Venzi space for $\left\lbrace0,0,1,1 \right\rbrace $, hence its curvature $2$-forms are recurrent,
%----------------------------------------------------------------------------
\item  its conformal curvature 2-forms are recurrent for the 1-form $\left\lbrace 0,\frac{1}{r},0,0\right\rbrace $,
%---------------------------------------------------------------------------
\item  its Ricci $1$-forms are recurrent for the 1-form $\left\lbrace 0,-\frac{1}{r},0,0\right\rbrace $, 

\item it is Chaki pseudosymmetric for the $1$-form  $\left\lbrace 0,-\frac{1}{r},0,0\right\rbrace $,

\item it is Chaki pseudo Ricci symmetric for the $1$-form  $\left\lbrace 0,-\frac{1}{r},0,0\right\rbrace $,

%---------------------------------------------------------------------------
\item its Riemann curvature can be decomposed  as $R=\frac{r^2}{2 (-1+\alpha^2) }S\wedge S$. Hence, it is an Ein$(2)$ spacetime with  $S^2 = \frac{(-1+\alpha^2)}{r^2}S$,
%-----------------------------------------------------------------------

%---------------------------------------------------------------------

\item it is a generalized quasi-Einstein spacetime for $\alpha=\frac{1}{2}(r^2+\sqrt{4+r^4})$,  $\beta=\frac{1}{2}(r^2-\sqrt{4+r^4})$, $\gamma=1$, $\Theta$= $\left\lbrace -\frac{\sqrt{(2+r^4+r^2\sqrt{4+r^4})}}{\sqrt{2}},1,0,0\right\rbrace $ and
$\Sigma$=$\left\lbrace \frac{(r^2-\sqrt{4+r^4})\sqrt{(2+r^4+r^2\sqrt{4+r^4})}}{2\sqrt{2}},0,0,0\right\rbrace $  and

%\item Generalized quasi-Einstein manifold in the sense of De and Ghosh for $\alpha=r^2$, $\beta=1$, $\gamma=-1$, $\Pi$=$\left\lbrace -r(r^2-1)^{1/2},1,0,0\right\rbrace $ 
%and $\phi$= $\left\lbrace r,-\frac{(r^2-1)^{1/2}}{r},0,0\right\rbrace $,
%\item Pseudo quasi-Einstein manifold for 
\item its Ricci tensor is compatible for the curvature  $R$, $C$, $K$, $W$ and $P$.
%-------------------------------------------------------------------------

%\item Chaki pseudosymmetry for the $1$-form  $\left\lbrace 0,-\frac{l}{r},0,0\right\rbrace $ of Riemann and Ricci.
%item Weakly symmetric for $R$ and $S$ and hence pseudo Ricci symmetric.
\end{enumerate}
\end{thm}

\begin{cor}
	The PGM spacetime is Chaki pseudosymmetric and hence it is weakly symmetric in the sense of Tam\'{a}ssy and Binh  for the associated $1$-forms  $\Pi$=$\left\lbrace 0,-\frac{2}{r},0,0\right\rbrace $, $\Psi$=$\left\lbrace 0,-\frac{1}{r},0,0\right\rbrace $ and $\Phi$=$\left\lbrace 0,-\frac{1}{r},0,0\right\rbrace $.
\end{cor}
%

%%%%%%%%%%%%%%%%%%%%%%%%%%%%%%%%%%%%%%%%%%%%%%%%%%%%%%%%%%%%%%%%%%%%%%%%%%%%%%%%%%%%%%%%%%%%%%%%%%%%%%%%%%%%%%

\begin{rem}
	From the calculation with various tensors, it can be mentioned that the PGM spacetime \eqref{GMS} does not admit certain geometric structures, which are described as follows: 

\begin{enumerate}[label=(\roman*)]
	
	\item it is neither recurrent nor recurrent for $C$, $P$, $W$, $K$, 
%\item  not pseudosymmetric for $R\cdot R=f_RQ(g,R)$ for any smooth function $f_R$ and hence Ricci pseudosymmetric, conharmonic pseudosymmetric, concircular pseudosymmetric, conformal pseudosymmetric and projective pseudosymmetric, 
%\item not pseudosymmetric due to projective curvature tensor for $P\cdot R=f_RQ(g,R)$ and hence not   $P\cdot S=f_RQ(g,S)$,  $P\cdot C=f_RQ(g,C)$,  $P\cdot P=f_RQ(g,P)$,  $P\cdot W=f_RQ(g,W)$ and $P\cdot K=f_RQ(g,K)$,

\item its Ricci tensor is neither of Codazzi type nor cyclic parallel,
\item it is not a Venzi space for $C$, $P$, $W$, $K$,
%------------------------------------------------------------------
%\item not conharmonically recurrent and hence not recurrent for projectively, concircularly, 
%\item  generalized Roter type,
%------------------------------------------------------------------
\item it is not Ricci generalized pseudosymmetric  (i.e., $R\cdot R$ and $Q(S,R)$ are not linearly independent),
%\item not  Super generalized recurrent and hence not weakly generalized recurrent and hyper generalized recurrent,
\item it is neither Einstein nor quasi-Einstein and
%------------------------------------------------------------------
\item its curvature $2$-forms are not recurrent for   $K$, $P$ and $W$.
%\item  not chaki pseudosymmetric for $C$, $K$, $P$ and $W$. 
%-----------------------------------------------------------------
%----------------------------------------------------------------
\end{enumerate}
\end{rem}
%
%%%%%%%%%%%%%%%%%%%%%%%%%%%%%%%%%%%%%%%%%%%%%%%%%%%%%%%%%%%%%%%%%%
%
%%%%%%%%%%%%%%%%%%%%%%%%%%%%%%%%%%%%%%%%%%%%%%%%%%%%%%%%%%%%%%%%%%

%%%%%%%%%%%%%%%%%%%%%%%%%%%%%%%%%%%%%%%%%%%%%%%%%%%%%%%%%%%%%%%%%%%%%%%%%%%%%%%%%%%%%%%%%%%%%%%%%%%%%%%%%%%%%%

%%%%%%%%%%%%%%%%%%%%%%%%%%%%%%%%%%%%%%%%%%%%%%%%%%%%%%%%%%%%%%%%%%%%%%%%%%%%%%%%%%%%%%%%%%%%%%%%%%%%%%%%%%%%%%%%%%%%%%%%%%%%%%%%%%%%%%%%%%%%%%%%%%%%%%%%
\section{\bf Pseudosymmetric structure admitted by the Energy-momentum tensor of PGM  spacetime}
%============================================================
From the EFEs,  the stress energy-momentum tensor $T$  of a spacetime is given by
$$ T=\frac{1}{\tau}\Big(S-\frac{\kappa}{2}\,g\Big),$$
where $\tau=\frac{8\pi G}{c^4}$, $c$ is the velocity of light in vacuum, and $G$ is the gravitational constant. The only non-vanishing components (upto symmetry) of the energy momentum tensor $T$ are given by 
$$\begin{array}{c}
T_{11}=\frac{-1+\alpha^2}{8 r^2}, \ \ T_{22}=\frac{(1-\alpha^2)}{8 r^2 \alpha^2}.
\end{array}$$
\indent Hence the non-vanishing components of covariant derivative of the energy momentum tensor  $T$ are given by 
$$\begin{array}{c}
T_{11,2}=-\frac{(-1+\alpha^2)}{4 r^3}, \ \ T_{22,2}=\frac{(-1+\alpha^2)}{4 r^3 \alpha^2}, \\ T_{23,3}=-\frac{(-1+\alpha^2)}{8 r}, \ \ T_{24,4}=-\frac{(-1+\alpha^2) \sin^2 \theta}{8 r}.
\end{array}$$

Let $C\cdot T=\mathcal{C}^1$, $W\cdot T=\mathcal{W}^1$, $K\cdot T=\mathcal{K}^1$ and  $Q(g,T)= \mathcal{Q}^1$.
Then the components other than zero   of the tensor $\mathcal{C}^1$, $\mathcal{W}^1$, $\mathcal{K}^1$ and $\mathcal{Q}^1$ are calculated as follows:
$$\begin{array}{c}
\mathcal{C}^1_{1313}=\frac{(-1+\alpha^2)^2}{48 r^2}=\frac{1}{\sin^2\theta} \mathcal{C}^1_{1414}, \ \  \mathcal{C}^1_{2323}=-\frac{(-1+\alpha^2)^2}{48 r^2 \alpha^2}=\frac{1}{\sin^2\theta} \mathcal{C}^1_{2424};
\end{array}$$
$$\begin{array}{c}
\mathcal{W}^1_{1313}=\frac{(-1+\alpha^2)^2}{48 r^2}=\frac{1}{\sin^2\theta} \mathcal{W}^1_{1414}, \ \  \mathcal{W}^1_{2323}=-\frac{(-1+\alpha^2)^2}{48 r^2 \alpha^2}=\frac{1}{\sin^2\theta} \mathcal{W}^1_{2424};
\end{array}$$
$$\begin{array}{c}
\mathcal{K}^1_{1313}=\frac{(-1+\alpha^2)^2}{16 r^2}=\frac{1}{\sin^2\theta} \mathcal{K}^1_{1414}, \ \  \mathcal{K}^1_{2323}=-\frac{(-1+\alpha^2)^2}{16 r^2 \alpha^2}=\frac{1}{\sin^2\theta} \mathcal{K}^1_{2424};
\end{array}$$
$$\begin{array}{c}
\mathcal{Q}^1_{1313}=-\frac{(-1+\alpha^2)}{8}=\frac{1}{\sin^2\theta} \mathcal{Q}^1_{1414}, \ \ 
\mathcal{Q}^1_{2323}=\frac{(-1+\alpha^2)}{8 \alpha^2}=\frac{1}{\sin^2\theta} \mathcal{Q}^1_{2424}.
\end{array}$$

From the above calculations we can state the following:
\begin{thm}\label{mainthm2}
	The PGM spacetime \eqref{GMS} admits certain pseudosymmetric type curvature conditions for the energy momentum tensor $T$  given as follows:
	\begin{enumerate}[label=(\roman*)]
		
		\item $C\cdot T=-\frac{(-1+\alpha^2)}{6 r^2}Q(g,T)$,  i.e., the nature of the energy momentum tensor is conformally pseudosymmetric,
		%-------------------------------------------------------
		\item $W\cdot T=-\frac{(-1+\alpha^2)}{6 r^2}Q(g,T)$,  i.e., the nature of the energy momentum tensor is concircularly pseudosymmetric,
		%----------------------------------------------------- 
		\item $K\cdot T=-\frac{(-1+\alpha^2)}{2 r^2}Q(g,T)$,  i.e., the nature of the energy momentum tensor is conharmonically pseudosymmetric,
		%---------------------------------------------------
		%\item $P\cdot R=\frac{1}{6}Q(T,R)$ and also
		%-------------------------------------------------
		\item the energy momentum tensor $T$ is compatible for Riemann,  projective, conharmonic, conformal and concircular curvature tensors.
	\end{enumerate}
\end{thm}

\section{\bf Curvature collineation and inheritance realized by  PGM spacetime}

Let $\chi(M)$ be the Lie algebra of all smooth vector fields on an $n$-dimensional smooth semi-Riemannian manifold $ M $. Then the Lie subalgebra $\mathcal K(M)$ of all Killing vector fields contains at most $n(n+1)\slash2$ linearly independent Killing vector fields. If $M$ is of constant scalar curvature, $\mathcal K(M)$ consists of exactly $n(n+1)\slash2$ linearly independent vector fields. In section \ref{com}, it has shown that the PGM spacetime possesses non-constant scalar curvature $\kappa$ given by $\frac{2\,(\alpha^2-1)}{r^2}$. In this section, some Killing vector fields on PGM spacetime will exhibit, and we show that the PGM spacetime admits curvature collineation, Ricci collineation and curvature inheritance for some non-Killing vector fields.

\begin{pr}
	The PGM spacetime admits motion for the vector fields $\frac{\partial}{\partial t}$ and $\frac{\partial}{\partial \phi}$, i.e., the vector fields $\frac{\partial}{\partial t}$ and $\frac{\partial}{\partial \phi}$ on  PGM spacetime are Killing  ($\pounds_{\frac{\partial}{\partial t}}g=0$ and $\pounds_{\frac{\partial}{\partial \phi}}g=0$).
\end{pr}

\begin{cor}
	As $\frac{\partial}{\partial t}$ and $\frac{\partial}{\partial \phi}$ are Killing vector fields, the vector field  $\lambda\frac{\partial}{\partial t}+\mu\frac{\partial}{\partial \phi}$ is also Killing for any constants $\lambda$ and $\mu$, i.e.,  $\pounds_{\lambda\frac{\partial}{\partial t}+\mu\frac{\partial}{\partial \phi}}g=0$ for all real numbers $\lambda$ and $\mu$.
\end{cor}
In this section we have considered the non-Killing vector fields $\frac{\partial}{\partial r}$, $\frac{\partial}{\partial \theta}$, $\lambda\frac{\partial}{\partial r}+\mu\frac{\partial}{\partial \theta}$ ($\lambda$ and $\mu$ are constants), in the direction of which the Lie derivative of various tensors are computed.

The non-zero components of the (1,3)-type curvature tensor $\widetilde R$ are given as follows:
\begin{equation}\label{R13component}
	 \widetilde R^3_{434}=(1-\alpha^2)\sin^2\theta, \ \ \ 
	\widetilde R^3_{334}=-(1-\alpha^2)
\end{equation}
and the non-vanishing components of the (0,4)-type curvature tensor $R$ are given in \eqref{R04}. From the components of $\widetilde{R}$ provided in \eqref{R13component}, we have $\pounds_{\frac{\partial}{\partial r}} \widetilde R=0$, which leads to the following:

\begin{pr}
		The PGM spacetime admits curvature collineation for the non-Killing vector field $\xi=\frac{\partial}{\partial r}$ as it possesses $\pounds_\xi \widetilde R=0$.
\end{pr}
 Again, Duggal  (Theorem 3, \cite{Duggal1992}) proved that if a manifold admits curvature inheritance, it also realizes Ricci inheritance, and hence the above proposition implies the following: 

\begin{cor}
	The PGM spacetime realizes Ricci collineation  for the non-Killing vector field $\xi=\frac{\partial}{\partial r}$, i.e., $\pounds_\xi S=0$.
\end{cor}

For the non-Killing vector field  $\eta=\frac{\partial}{\partial \theta}$, the non-vanishing components of $\pounds_\eta \widetilde R$ and $\pounds_\eta S$,  are computed as follows:
\begin{equation}\label{LvR13}
\begin{array}{c}
	(\pounds_\eta \widetilde R)^3_{434}=
	(1-\alpha^2) \sin2\theta=
	-(\pounds_\eta \widetilde R)^3_{443},
\end{array}
\end{equation}
\begin{equation}\label{LvS}
	\begin{array}{c}
	(\pounds_\eta S)_{44}=
	-(1-\alpha^2) \sin2\theta.
\end{array}
\end{equation}
From the tensor components in \eqref{LvR13} and \eqref{LvS}, we note the following remarks:
\begin{rem}\label{rm1}
	For the non-Killing vector field $\eta=\frac{\partial}{\partial \theta}$, there exists no scalar function $\lambda$ such that the PGM spacetime possesses the relation $\pounds_\eta \widetilde R=\lambda \widetilde R$, i.e., with respect to the non-Killing vector field $\frac{\partial}{\partial \theta}$ the PGM spacetime admits neither curvature collineation  nor curvature inheritance (in sense of Definition \ref{def_CI}) for the (1,3)-type curvature tensor $\widetilde{R}$.
\end{rem}

\begin{rem}\label{rm2}
	For the non-Killing vector field $\eta=\frac{\partial}{\partial \theta}$, there exists no scalar function $\lambda$ such that the PGM spacetime realizes  $\pounds_\eta S=\lambda S$, i.e.,  with respect to the non-Killing vector field $\frac{\partial}{\partial \theta}$ the PGM spacetime possesses neither Ricci collineation nor Ricci inheritance.
\end{rem}

Now, for the non-Killing vector fields $\xi=\frac{\partial}{\partial r}$ and $\eta=\frac{\partial}{\partial \theta}$, the non-vanishing components of $\pounds_\xi R$ and $\pounds_\eta R$,  are calculated as follows:
$$\begin{array}{c}
	(\pounds_\xi R)_{3434}=
	(\pounds_\xi R)_{4343}=
	-2(\alpha^2-1)r \sin^2\theta=
	-(\pounds_\xi R)_{3443}=
	-(\pounds_\xi R)_{4334},
\end{array}$$
$$\begin{array}{c}
	(\pounds_\eta R)_{3434}=
	(\pounds_\eta R)_{4343}=
	-(\alpha^2-1)r^2 \sin2\theta=
	-(\pounds_\eta R)_{3443}=
	-(\pounds_\eta R)_{4334}.
\end{array}$$

This leads to the following:
\begin{pr}
	The PGM spacetime admits curvature inheritance (in the sense of Definition \ref{def_GCI}) for the vector fields $\xi=\frac{\partial}{\partial r}$ and $\eta=\frac{\partial}{\partial \theta}$ as it realizes the relations
	$$\pounds_\xi R=\frac{2}{r}R   \ \ \text{ and }\ \  \pounds_\eta R=2 \cot\theta \ R .$$
\end{pr}

If $\lambda$ and $\mu$ are any non-zero constants, the non-zero components of $ \pounds_\xi R $ for the vector field $V={\lambda\frac{\partial}{\partial r}+\mu\frac{\partial}{\partial \theta}}$ are given as follows:
$$\begin{array}{c}
	(\pounds_V R)_{3434}=
	(\pounds_V R)_{4343}=
	-2(\alpha^2-1)r \sin\theta(\mu r \cos \theta+\lambda \sin\theta)=
	-(\pounds_V R)_{3443}=
	-(\pounds_V R)_{4334}.
\end{array}$$

The above components of $\pounds_{{\lambda\frac{\partial}{\partial r}+\mu\frac{\partial}{\partial \theta}}} R$ lead to the following:
\begin{pr}
	For the vector field $\xi={\lambda\frac{\partial}{\partial r}+\mu\frac{\partial}{\partial \theta}}$, the PGM spacetime possesses curvature inheritance (Definition \ref{def_GCI}) in the sense of Shaikh and Datta \cite{ShaikhDatta2022} as it satisfies the relation  
	$$\pounds_\xi R=\frac{2(\lambda+\mu r \cot \theta)}{r}R.$$
\end{pr}
Incorporating the above propositions and their consequences, we can state the following:

\begin{thm} \label{th_51}
	The PGM spacetime reveals the following symmetry properties:
	\begin{enumerate} [label=(\roman*)]
		\item it admits motion for the vector fields $\frac{\partial}{\partial t}$ and $\frac{\partial}{\partial \phi}$,
		
		\item if $\lambda,\mu$ are any non-zero constants, it possesses motion for the vector field $\lambda\frac{\partial}{\partial t}+\mu\frac{\partial}{\partial \phi}$,
		
		\item it admits curvature collineation (in the sense of Definition \ref{def_CI}) and hence Ricci collineation with respect to the non-Killing vector field $\frac{\partial}{\partial r}$, in fact,  $\pounds_{\frac{\partial}{\partial r}} \widetilde R=0$ and $\pounds_{\frac{\partial}{\partial r}} S=0$,

		\item it admits curvature inheritance (in the sense of Definition \ref{def_GCI}) for the non-Killing vector fields $\frac{\partial}{\partial r}$ and $\frac{\partial}{\partial \theta}$, in fact, $$\pounds_{\frac{\partial}{\partial r}} R=\frac{2}{r}R   \ \ \text{ and }\ \  \pounds_{\frac{\partial}{\partial \theta}} R=2 \cot\theta \ R ,$$
		
		\item for any non-zero constants $\lambda,\mu$ it realizes curvature inheritance (in the sense of Definition \ref{def_GCI}) for the non-Killing vector field ${\lambda\frac{\partial}{\partial r}+\mu\frac{\partial}{\partial \theta}}$, in fact, 	$$\pounds_{\lambda\frac{\partial}{\partial r}+\mu\frac{\partial}{\partial \theta}} R=\frac{2(\lambda+\mu r \cot \theta)}{r}R.$$
		
	\end{enumerate}
		
\end{thm}

\begin{rem} \label{rm3}
	It is interesting to note that the PGM spacetime concerning the non-Killing vector field $\frac{\partial}{\partial r}$ admits curvature collineation for the (1,3)-type curvature tensor $\widetilde R$. But it does not realize curvature collineation for the (0,4)-type curvature tensor $R$, whereas it possesses curvature inheritance for the (0,4)-type curvature tensor $R$. Also, we show that for the non-Killing vector field $\frac{\partial}{\partial \theta}$, the PGM spacetime admits curvature inheritance for the (0,4)-type curvature tensor $R$, but it realizes neither curvature collineation nor curvature inheritance for the (1,3)-type curvature tensor $\widetilde{R}$. Hence it follows that the notion of curvature inheritance (resp., curvature collineation)  for (1,3)-type curvature tensor (Definition \ref{def_CI}) in the sense of Duggal \cite{Duggal1992} and the notion of curvature inheritance (resp., curvature collineation)  for (0,4)-type curvature tensor (Definition \ref{def_GCI}) in the sense of Shaikh {\it et al.} \cite{ShaikhDatta2022} are not equivalent.
\end{rem}

\section{\bf Conclusions}

In this paper, we have investigated various curvature-restricted  geometric properties of PGM spacetime. It has proved that this spacetime is not Ricci generalized pseudosymmetric but admits various types of pseudosymmetric type curvature conditions, such as pseudosymmetry due to Weyl conformal curvature tensor, pseudosymmetry due to conharmonic curvature tensor, and Ricci generalized conformal pseudosymmetry due to projective curvature tensor. Also, it has proved that such a spacetime is Einstein manifold of degree $2$,  generalized quasi-Einstein and $2$-quasi Einstein  manifold (see, Theorem~\ref{mainthm}). Moreover, the energy-momentum tensor of the spacetime satisfies several pseudosymmetric type curvature conditions, and both its Ricci tensor and energy-momentum tensor are compatible with Riemann, conformal, projective, conharmonic and concircular curvature (see, Theorem~\ref{mainthm2}). Finally, it has shown that the PGM spacetime admits curvature collineation, Ricci collineation for the (1,3)-curvature tensor and curvature inheritance for the (0,4) curvature tensor concerning the non-Killing vector fields (see, Theorem~\ref{th_51}). Also, a few non-Killing vector fields have exhibited (see, Remark  \ref{rm3}), for which it has shown that the notions of curvature inheritance (also, of curvature collineation) for the $(1,3)$-type curvature tensor by Duggal \cite{Duggal1992} and for the $(0,4)$-type curvature tensor by Shaikh {\it et al.} \cite{ShaikhDatta2022} are distinct (see, Remark \ref{rm1} and Remark \ref{rm3}).

\section*{Acknowledgment}

We sincerely acknowledged the anonymous referee for his/her valuable comments and suggestions. One of the author B. R. Datta is grateful to the Council of Scientific and Industrial Research (CSIR File No.: 09/025(0253)/2018-EMR-I), Govt. of India, for the award of SRF (Senior Research Fellowship). All the algebraic computations of {\it Sect. 3-5} are performed by a program in Wolfram Mathematica developed by the first author A. A. Shaikh.


\begin{thebibliography}{99}\baselineskip=16pt
	%============================================

\bibitem{Duggal1992} K. L. Duggal, %\emph{Curvature inheritance symmetry in Riemannian spaces with applications to fluid space times},
J. Math. Phys.  \textbf{33} (1992) 2989.

\bibitem{ShaikhDatta2022} A. A. Shaikh and B. R. Datta, %\emph{Ricci solitons and curvature inheritance on Robinson-Trautman spacetimes},  
https://doi.org/10.48550/arXiv.2209.03749.


\bibitem{Cart26} \'E. Cartan, %\emph{Sur une classe remarquable d'espaces de Riemannian}, 
Bull. Soc. Math. France, \textbf{54} (1926) 214.	

\bibitem{Cart46} \'E. Cartan, \emph{La g\' eom\' etrie des espaces de Riemann}, 1$^{st}$ Ed.,  Gauthier-Villars, Paris, \textbf{1925}, Cahiers scientifiques fascicule II.,  Gauthier-Villars, Paris, \textbf{1928}, \emph{Le\c cons sur la g\' eom\' etrie des espaces de Riemann}, 2$^{nd}$ Ed. rev. et. augm., Gauthier-Villars, Paris, \textbf{1946}.

\bibitem{Szab82} Z. I. Szab\'o, %\emph{Structure theorems on Riemannian spaces satisfying $ R(X, Y)\cdot R = 0$, I. The local version}, 
J. Diff. Geom. \textbf{17} (1982) 531--582.

\bibitem{Szab84} Z. I. Szab\'o, %\emph{Classification and construction of complete hypersurfaces satisfying  $R(X, Y)\cdot R = 0$}, 
Acta Sci. Math. \textbf{47} (1984) 321--348.

\bibitem{Szab85} Z. I. Szab\'o, %\emph{Structure theorems on Riemannian spaces satisfying $R(X, Y)\cdot R = 0$, II, The global version}, 
Geom. Dedicata \textbf{19} (1985) 65--108.	

\bibitem{Chak87} M. C. Chaki, %\emph{On pseudosymmetric manifolds}, 
An. \c{S}tiin\c{t}.  Univ. AL. I. Cuza Ia\c{s}i. Mat. (N.S.)  Sect. Ia, \textbf{33(1)} (1987), 53--58.

\bibitem{Chak88} M. C. Chaki, %\emph{On pseudo Ricci symmetric manifolds}, 
Bulgarian J. Phys. {\bf 15} (1988) 526--531.
	
\bibitem{AD83} A. Adam\'{o}w and R. Deszcz, %\emph{On totally umbilical submanifolds of some class of Riemannian manifolds},
Demonstratio Math., \textbf{16} (1983) 39--59.

\bibitem{TB89} L. T$\acute{\mbox{a}}$massy and T. Q. Binh, %\emph{On weakly symmetric and weakly projective symmetric Riemannian manifolds}, 
Colloq. Math. Soc. J. Bolyai \textbf{50} (1989) 663--670.

\bibitem{TB93} L. Tam\'{a}ssy and T. Q. Binh, %\emph{On weak symmetries of  Einstein and Sasakian manifolds},
Tensor (N. S.) \textbf{53} (1993) 140--148.
	
\bibitem{Ruse46} H. S. Ruse, %\emph{On simply harmonic spaces},
J. London Math. Soc. \textbf{21} (1946) 243--247.

\bibitem{Ruse49a} H. S. Ruse, %\emph{On simply harmonic `kappa spaces' of four dimensions},
Proc. London Math. Soc. \textbf{50} (1949) 317--329.

\bibitem{Ruse49b} H. S. Ruse, %\emph{Three dimensional spaces of recurrent curvature},
Proc. London Math. Soc. \textbf{50} (1949) 438--446.	

\bibitem{Walk50} A. G. Walker, %\emph{On Ruse's spaces of recurrent curvature}, 
Proc. London Math. Soc. \textbf{52} (1950) 36--64.

\bibitem{Bess87} A. L. Besse, \emph{Einstein Manifolds}, Springer-Verlag, Berlin, Heidelberg (1987).

\bibitem{LR89} D. Lovelock and H. Rund, \emph{Tensors, differential forms and variational principles}, Courier Dover ublications (1989).

\bibitem{SAR13} A. A. Shaikh, F. R. Al-Solamy and I. Roy, %\emph{On the existence of a new class of semi-Riemannian manifolds},
Math. Sci. \textbf{7} (2013) 46.

\bibitem{SP10} A. A. Shaikh and A. Patra, %\emph{On a generalized class of recurrent manifolds},
Arch. Math. (BRNO) {\bf 46} (2010) 71--78.

\bibitem{SR10} A. A. Shaikh and I. Roy, %\emph{On quasi generalized recurrent manifolds},
Math. Pannon \textbf{21(2)} (2010) 251--263.

\bibitem{SR11} A. A. Shaikh and I. Roy, %\emph{On weakly generalized recurrent manifolds}, Ann. Univ. Sci. Budapest, 
 E$\ddot{\mbox{o}}$tv$\ddot{\mbox{o}}$s Sect. Math. \textbf{54} (2011) 35--45.

\bibitem{SRK15} A. A. Shaikh, I. Roy and H. Kundu, %\emph{On the existence of a generalized class of recurrent manifolds},
An. \c{S}tiin\c{t}. Univ. Al. I. Cuza Ia\c{s}i. Mat. (N. S.), \textbf{LXIV(2)} (2018), pp. 233--251.

\bibitem{SRK17} A. A. Shaikh, I. Roy and H. Kundu, %\emph{On some generalized recurrent manifolds},
Bull. Iranian Math. Soc. \textbf{43(5)} (2017) 1209--1225.

\bibitem{SK16} A. A. Shaikh and H. Kundu, %\emph{On warped product generalized Roter type manifolds},
Balkan J. Geom. Appl. \textbf{21(2)} (2016) 82--95.

\bibitem{ADEHM14} K. Arslan, R. Deszcz, R. Ezenta\c{s}, M. Hotlo\'{s} and C. Murathan, %\emph{On generalized Robertson-Walker spacetimes satisfying some curvature condition}, 
Turkish J. Math. \textbf{38(2)} (2014) 353--373.

\bibitem{DK99} R. Deszcz and M. Kucharski, %\emph{On curvature properties of certain generalized Robertson-Walker spacetimes},
Tsukuba J. Math. \textbf{23(1)} (1999) 113--130.

\bibitem{DHJKS14} R. Deszcz, M. Hotlo\'{s}, J. Je\l owicki, H. Kundu and A. A. Shaikh, %\emph{Curvature properties of G\"{o}del metric},
Int. J. Geom. Methods Mod. Phys. \textbf{11} (2014) 1450025. Erratum: %\emph{Curvature properties of Gödel metric},
Int. J. Geom. Methods Mod. Phys. \textbf{16} (2019) 1992002.

\bibitem{SDKC19} A. A. Shaikh, L. Das, H. Kundu and D. Chakraborty, %\emph{Curvature properties of Siklos metric},
Diff. Goem.-Dyn. Syst. \textbf{21} (2019) 167--180.

\bibitem{SAA18} A. A. Shaikh, M. Ali and Z. Ahsan, %\emph{Curvature properties of Robinson-Trautman metric},
J. Geom. \textbf{109 (2)} (2018) 1.

\bibitem{Kowa06} D. Kowalczyk, %\emph{On the Reissner-Nordström-de Sitter type spacetimes},
Tsukuba J. Math. \textbf{30(2)} (2006) 363.

\bibitem{KLD1969} G. H. Katzin, J. Livine and W. R. Davis, %\emph{Curvature collineations: A fundamental symmetry property of the space-times of general relativity defined by the vanishing Lie derivative of the Riemann curvature tensor},
J. Math. Phys., \textbf{10(4)} (1969) 617.

\bibitem{KLD1970} G. H. Katzin, J. Livine and W. R. Davis, %\emph{Groups of curvature collineations in Riemannian space-times which admit fields of parallel vectors}, 
J. Math. Phys. \textbf{11} (1970) 1578.

\bibitem{Ahsan1977_1055} Z. Ahsan, %\emph{Algebra of space-matter tensor in general relativity},
Indian J. Pure Appl. Math. \textbf{8} (1977) 1055.
	
\bibitem{Ahsan1987} Z. Ahsan, %\emph{On the Nijenhuis tensor for null electromagnetic field},
J. Math. Phys. Sci. \textbf{21(5)} (1987) 515.
	
\bibitem{Ahsan1995} Z. Ahsan, %\emph{Symmetries of the Electromagnetic Fields in General Relativity},
Acta Phys. Sinica \textbf{4} (1995) 337.
	
\bibitem{Ahsan1996} Z. Ahsan, %\emph{A symmetry property of the space-time of general relativity in terms of the space-matter tensor},
Braz. J. Phys. \textbf{26} (1996) 572.
	
\bibitem{Ahasan2005} Z. Ahsan, %\emph{On a geometrical symmetry of the space-time of General Relativity},
Bull. Call. Math. Soc. \textbf{97} (2005) 191.
		
\bibitem{AhsanAli2014} Z. Ahsan and M. Ali, %\emph{On some properties of $W$-curvature tensor},
Palestine J. Math. \textbf{3(1)} (2014) 61.
	
\bibitem{AA2012} Z. Ahsan and M. Ali, %\emph{Symmetries of Type D Pure Radiation Fields},
Int. J. Theor. Phys. \textbf{51} (2012) 2044.
	
\bibitem{AH1980} Z. Ahsan and S. I. Husain, %\emph{Null electromagnetic fields, total gravitational radiation and collineations in general relativity}, 
Annali di Mathematical Pura ed Applicata \textbf{126} (1980) 379396.
	
\bibitem{AliAhsan2012} M. Ali and Z. Ahsan, %\emph{Ricci Solitons and Symmetries of Spacetime Manifold of General Relativity},
Global J. Adv. Research Classical Mod. Geom. \textbf{1} (2012) 75.
	
\bibitem{SASZ2022} A. A. Shaikh, M. Ali, M. Salman and F. O. Zengin, %\emph{Curvature inheritance symmetry on M-projectively flat spacetimes},
Int. J. Geom. Methods Mod. Phys. {\bf 20}, 2350088 (2023).
	
\bibitem{Ahsan1978} Z. Ahsan, %\emph{Collineation in electromagnetic field in general relativity- The null field case},
Tamkang J. Maths. \textbf{9} (1978) 237.
	
\bibitem{Ahsan1977_231} Z. Ahsan, %\emph{Algebraic classification of space-matter tensor in general relativity},
Indian J. Pure Appl. Math. \textbf{8} (1977) 231.

\bibitem{TWBK} T. W. B. Kibble, %\emph{Some implications of a cosmological phase transition},
Phys. Rep. {\bf 67}, 183 (1980).
	
\bibitem{AVV} A. Vilenkin, %\emph{Cosmic strings and domain walls},
Phys. Rep. {\bf 121}, 263 (1985).
	
\bibitem{WAH} W. A. Hiscock, %\emph{Exact gravitational field of a string},
Phys. Rev. {\bf D 31}, 3288 (1985). 
	
\bibitem{MBAV} M. Barriola and A. Vilenkin, %\emph{Gravitational field of a global monopole},
Phys. Rev. Lett. {\bf 63}, 341 (1989).

\bibitem{ERFM2} E. R. Bezerra de Mello, %\emph{Physics in the Global Monopole Spacetime},
Braz. J. Phys. {\bf 31}, 211 (2001).

\bibitem{ae} A. Edery and Y. Nakayama, %\emph{Gravitating magnetic monopole via the spontaneous symmetry breaking of pure $\mathcal{R}^2$ gravity},
Phys. Rev. {\bf D 98}, 064011 (2018).

\bibitem{ERBM} E. R. Bezerra de Mello and C. Furtado, %\emph{Nonrelativistic scattering problem by a global monopole},
Phys. Rev. {\bf D 56}, 1345 (1997).
 
\bibitem{ALCO2} A. L. C. de Oliveira and E. R. B. de Mello, %\emph{Nonrelativistic Charged Particle-Magnetic Monopole Scattering in the Global Monopole Background}, 
Int. J. Mod. Phys. {\bf A 18}, 3175 (2003).
	
\bibitem{ERBM44} A. L. Cavalcanti de Oliveira and E. R. B. de Mello, %\emph{Nonrelativistic Scattering Analysis of Charged Particle by a Magnetic Monopole in the Global Monopole Background}, 
Int. J. Mod. Phys. {\bf A 18}, 2051 (2003).

\bibitem{AAS} E. R. Bezerra de Mello, A. A. Saharian, %\emph{Scalar self-energy for a charged particle in global monopole spacetime with a spherical boundary}, 
Class. Quantum Grav. {\bf 29}, 135007 (2012).
	
\bibitem{AAS2} E. R. Bezerra de Mello and A. A. Saharian, %\emph{Electrostatic self-interaction in the spacetime of a global monopole with finite core}, 
Class. Quantum Grav. {\bf 24}, 2389 (2007).
	
\bibitem{AAS3} D. Barbosa, U. de Freitas and E. R. Bezerra de Mello, %\emph{Induced self-energy on a static scalar charged particle in the spacetime of a global monopole with finite core}, 
Class. Quantum Grav. {\bf 28}, 065009 (2011).
	
\bibitem{AAS4} F. C. Carvalho and E. R. Bezerra de Mello, %\emph{Vacuum polarization for a massless scalar field in the global monopole spacetime at finite temperature}, 
Class. Quantum Grav. {\bf 18}, 1637 (2001).
	
\bibitem{AAS5} F. C. Carvalho and E. R. B. de Mello, %\emph{Vacuum polarization for a massless spin-1/2 field in the global monopole spacetime at nonzero temperature},
Class. Quantum Grav. {\bf 18}, 5455 (2001).
	
\bibitem{AAS6} E. R. B. de Mello, %\emph{Vacuum polarization effects in the global monopole spacetime in the presence of the Wu-Yang magnetic monopole}, 
Class. Quantum Grav. {\bf 19}, 5141 (2002).

\bibitem{ALCO} A. L. Cavalcanti de Oliveira and E. R. Bezerra de Mello, %\emph{Exact solutions of the Klein–Gordon equation in the presence of a dyon, magnetic flux and scalar potential in the spacetime of gravitational defects},
Class. Quantum Grav. {\bf 23}, 5249 (2006).

\bibitem{ABHA} A. Boumali and H. Aounallah, %\emph{Exact Solutions of Scalar Bosons in the Presence of the Aharonov-Bohm and Coulomb Potentials in the Gravitational Field of Topological Defects},
Adv. High Energy Phys. {\bf 2018}, 1031763 (2018). 

\bibitem{EAFB} E. A. F. Braganca, R. L. L. Vit\'{o}ria, H. Belich and E. R. Bezerra de Mello, %\emph{Relativistic quantum oscillators in the global monopole spacetime},
Eur. Phys. J. C {\bf 80}, 206 (2020).

\bibitem{SZ} M. de Montigny, H. Hassanabadi, J. Pinfold and S. Zare, %\emph{Exact solutions of the generalized Klein–Gordon oscillator in a global monopole space-time},
Eur. Phys. J. Plus {\bf 136}, 788 (2021).
 
\bibitem{SR2} F. Ahmed, %\emph{Relativistic motions of spin-zero quantum oscillator field in a global monopole space-time with external potential and AB-effect},
Sci. Rep. {\bf 12}, 8794 (2021).

\bibitem{SZ3} M. de Montigny, J. Pinfold, S. Zare and H. Hassanabadi, %\emph{Klein-Gordon oscillator in a global monopole space–time with rainbow gravity}, 
Eur. Phys. J. Plus {\bf 137}, 54 (2022).

\bibitem{CF44} C. Furtado and F. Moraes, %\emph{Harmonic oscillator interacting with conical singularities},
J. Phys. A: Math. Gen. {\bf 33}, 5513 (2000).
	
\bibitem{RV} R. L. L. Vitoria and H. Belich, %\emph{Harmonic oscillator in an environment with a pointlike defect},
Phys. Scr. {\bf 94}, 125301 (2019).
	
\bibitem{GAM} G. de A. Marques and V. B. Bezerra, %\emph{Non-relativistic quantum systems on topological defects space-times},
Class. Quantum Gravit. {\bf 19}, 985 (2002).
	
\bibitem{PN} P. Nwabuzor, C. Edet, A. N. Ikot, U. Okorie, M. Ramantswana, R. Horchani, A. H. A.-Aty, G. Rampho, %\emph{Analyzing the Effects of Topological Defect (TD) on the Energy Spectra and Thermal Properties of LiH, TiC and $I_2$ Diatomic Molecules},
Entropy {\bf 23}, 1060 (2021).

\bibitem{FA} F. Ahmed, %\emph{Approximate eigenvalue solutions with diatomic molecular potential under topological defects and Aharonov-Bohm flux field: application for some known potentials},
Mol. Phys. {\bf 120}, e2124935 (2022).
 
\bibitem{FA2} F. Ahmed, %\emph{Topological effects produced by point-like global monopole with Hulthen plus screened Kratzer potential on Eigenvalue solutions and NU-method},
Phys. Scr. {\bf 98}, 015403 (2023).

\bibitem{FA3} F. Ahmed, %\emph{Point-like defect on Schrödinger particles under flux field with harmonic oscillator plus Mie-type potential: application to molecular potentials},
Proc. R. Soc. A {\bf 479}, 20220624 (2023).

\bibitem{FA4} F. Ahmed, %\emph{Topological effects on non-relativistic eigenvalue solutions under AB-flux field with pseudoharmonic- and Mie-type potentials},
Commun. Theor. Phys. {\bf 75}, 055103 (2023).

\bibitem{FA5} F. Ahmed, %\emph{Topological defects with generalized Hulthen–Coulomb-inverse quadratic Yukawa potential on eigenvalue solution under Aharonov–Bohm flux field},
Int. J. Geom. Meth. Mod. Phys. {\bf 20}, 2350060 (2023).

\bibitem{FA6} F. Ahmed, %\emph{Radial solution of Schr\"{o}dinger equation with generalized inverse Hulthen and Yukawa potentials in topological defect}, 
EPL {\bf 141}, 25003 (2023).

\bibitem{FA7} F. Ahmed, %\emph{Hellmann potential and topological effects on non-relativistic particles confined by Aharonov–Bohm flux field},
Mol. Phys. {\bf 121}, e2155596 (2023).

\bibitem{FA8} F. Ahmed, %\emph{Quantum effects with Kratzer plus generalised Yukawa potential in a point-like global monopole using different approximation schemes},
Mol. Phys. {\bf 121}, e2198617 (2023).

\bibitem{FA9} F. Ahmed, %\emph{Radial solution of Schr\"{o}dinger equation with Hulthen-Yukawa-Inverse quadratic potential in a Point-Like defect under AB-flux field},
Rev. Mex. Fis. {\bf 69}, 030401 (2023).

\bibitem{FA10} F. Ahmed, %\emph{Eigenvalue spectra of non-relativistic particles confined by AB-flux field with Eckart plus class of Yukawa potential in point-like global monopole},
Indian J Phys. {\bf 97}, 2307 (2023).  


\bibitem{DGHS11} R. Deszcz, M. G\l ogowska, M. Hotlo\'s and K. Sawicz, \emph{A Survey on Generalized Einstein Metric Conditions}, Advances in Lorentzian Geometry, Proceedings of the Lorentzian Geometry Conference in Berlin, AMS/IP Studies in Advanced Mathematics, \textbf{49}, S.-T. Yau (series ed.), M. Plaue, A.D. Rendall and M. Scherfner (eds.), 2011, pp. 27-46.

\bibitem{Glog02} M. G\l ogowska, %\emph{Semi-Riemannian manifolds whose Weyl tensor is a Kulkarni-Nomizu square},
Publ. Inst. Math. (Beograd) (N.S.), \textbf{72(86)} (2002) 95.

\bibitem{SKA18} A. A. Shaikh, H. Kundu and Md. S. Ali, %\emph{On warped product super generalized recurrent manifolds},
An. \c{S}tiin\c{t}. Univ. Al. I. Cuza Ia\c{s}i. Mat. (N. S.), \textbf{LXIV(1)} (2018), pp. 85--99.

\bibitem{SC21} A. A. Shaikh and D. Chakraborty, %\emph{Curvature properties of Kantowski-Sachs metric},
J. Geom. Phys. \textbf{160} (2021) 103970.

\bibitem{SK19} A. A. Shaikh and H. Kundu, %\emph{On generalized Roter type manifolds},
Kragujevac J. Math. \textbf{43(3)} (2019) 471.

\bibitem{DG02} R. Deszcz and M. G\l ogowska, %\emph{Some examples of nonsemisymmetric Ricci-semisymmetric hypersurfaces},
Colloq. Math. \textbf{94} (2002) 87.

\bibitem{DGHS98} R. Deszcz, M. G\l ogowska, M. Hotlo\'{s} and Z. \d Sent\"{u}rk, %\emph{On certain quasi-Einstein semi-symmetric hypersurfaces},
Ann. Univ. Sci. Budapest E\"{o}tv\"{o}s Sect. Math. \textbf{41} (1998) 151.

\bibitem{DH03} R. Deszcz and M. Hotlo\'{s}, %\emph{On hypersurfaces with type number two in spaces of constant curvature},
Ann. Univ. Sci. Budapest E\"{o}tv\"{o}s Sect. Math. \textbf{46} (2003) 19.

\bibitem{SDHJK15} A. A. Shaikh, R. Deszcz, M. Hotlo\'{s}, J. Je\l owicki and H. Kundu, %\emph{On pseudosymmetric manifolds},
Publ. Math. Debrecen, \textbf{86(3-4)} (2015) 433.

\bibitem{DGPSS11} R. Deszcz, M. G\l ogowska, M. Plaue, K. Sawicz and M. Scherfner, %\emph{On hypersurfaces in space forms satisfying particular curvature conditions of Tachibana type},
Kragujevac J. Math. \textbf{35} (2011) 223.

\bibitem{SK14} A. A. Shaikh and H. Kundu, %\emph{On equivalency of various geometric structures},
J. Geom. \textbf{105} (2014) 139.

\bibitem{Tach74} S. Tachibana, %\emph{A Theorem on Riemannian manifolds of positive curvature operator},
Proc. Japan Acad. \textbf{50} (1974) 301.

\bibitem{Desz92} R. Deszcz, %\emph{On pseudosymmetric spaces},
Bull. Belg. Math. Soc., Ser. \textbf{A 44} (1992) 1.

\bibitem{Desz93} R. Deszcz, %\emph{Curvature properties of a pseudosymmetric manifolds},
Colloq. Math. \textbf{62} (1993) 139.

\bibitem{DGHS00} R. Deszcz, M. Glogowska, M. Hotlo\'{s} and K. Sawicz, %\emph{Hypersurfaces in space forms satisfying a particular roter type equation}, 
https://doi.org/10.48550/arXiv.2211.06700.

\bibitem{DGHZ15} R. Deszcz, M. G\l ogowska, M. Hotlo\'s and G. Zafindratafa, %\emph{On some curvature conditions of pseudosymmetric type},
Period. Math. Hungarica \textbf{70(2)} (2015) 153.

\bibitem{DGHZ16} R. Deszcz, M. G\l ogowska, M. Hotlo\'s and G. Zafindratafa, %\emph{Hypersurfaces in space 	forms satisfying some curvature conditions},
J. Geom. Phys. \textbf{99} (2016) 218.

\bibitem{SK18} A. A. Shaikh and H. Kundu, %\emph{On some curvature restricted geometric structures for projective curvature tensor},
Int. J. Geom. Meths Mod. Phys. \textbf{15}, 1850157 (2018).

\bibitem{SKppsnw} A. A. Shaikh and H. Kundu, %\emph{On warped product manifolds satisfying some pseudosymmetric type conditions},
Diff. Geom.- Dyn. Syst. \textbf{19} (2017) 119.

\bibitem{S09} A. A. Shaikh, %\emph{On pseudo quasi-Einstein manifolds},
Period. Math. Hungarica \textbf{59(2)} (2009) 119.

\bibitem{SKH11} A. A. Shaikh, Y. H. Kim and S. K. Hui, %\emph{On Lorentzian quasi Einstein manifolds},
J. Korean Math. Soc. \textbf{48} (2011) 669. Erratum: %\emph{On Lorentzian quasi Einstein manifolds,}
J. Korean Math. Soc. \textbf{48(6)} (2011) 1327.

\bibitem{SYH09} A. A. Shaikh, D. W. Yoon, and S. K. Hui, %\emph{ On quasi-Einstein spacetimes},
Tsukuba J. Math. {\bf 33(2)} (2009) 305.

\bibitem{ECS22} S. Eyasmin, D. Chakraborty and M. Sarkar, %\emph{ Curvature properties of Morris-Thorne Wormhole metric},
J. Geom. Phys. \textbf{174} (2022) 104457.

\bibitem{EC21} S. Eyasmin and D. Chakraborty, %\emph{Curvature properties of (t-z)-type plane wave metric},
J. Geom. Phys. \textbf{160} (2021) 104004.

\bibitem{SAAC20N} A. A. Shaikh, A. Ali, A. H. Alkhaldi and D. Chakraborty, %\emph{Curvature properties of Nariai spacetimes},
Int. J. Geom. Meths Mod. Phys. \textbf{17}, 2050034 (2020).

\bibitem{SDC} A. A. Shaikh, B. R. Datta and D. Chakraborty, %\emph{On some curvature properties of Vaidya-Bonner metric},
Int. J. Geom. Meths. Phys. {\bf 18}, 2150205 (2021).

\bibitem{C01} M. C. Chaki, %\emph{On generalized quasi-Einstein manifolds}
Publ. Math. Debrecen, \textbf{58} (2001) 683.

\bibitem{SK16srs} A. A. Shaikh and H. Kundu, %\emph{On curvature properties of Som-Raychaudhuri spacetime},
J. Geom. \textbf{108(2)} (2016) 501.

\bibitem{SAAC20} A. A. Shaikh, A. H. Alkhaldi and D. Chakraborty, %\emph{Curvature properties of Melvin magnetic metric},
J. Geom. Phys. \textbf{150} (2020) 103593.

\bibitem{SSC19} A. A. Shaikh, S. K. Srivastava and D. Chakraborty, %\emph{Curvature properties of anisotropic scale invariant metrics},
Int. J. Geom. Meths. Mod. Phys. {\bf 16}, 1950086 (2019).
 
\bibitem{DGJPZ13} R. Deszcz, M. G\l ogowska, L. Je\l owicki, M. Petrovi\'{c}-Torga\u{s}ev and G. Zafindratafa, %\emph{On Riemann and Weyl compatible tensors},
Publ. Inst. Math. (Beograd) (N.S.), \textbf{94(108)} (2013), pp. 111--124.


\bibitem{DGJZ-2016} R. Deszcz, M. G\l ogowska, J. Je\l owicki and Z. Zafindratafa, %\emph{Curvature properties of some class of warped product manifolds},
Int. J. Geom. Methods Mod. Phys. \textbf{13}, 1550135 (2016).

\bibitem{DGP-TV-2015} R. Deszcz, M. G\l ogowska, M. Petrovi\'{c}-Torga\u{s}ev and L. Verstraelen, %\emph{Curvature properties of some class of minimal hypersurfaces in Euclidean spaces},
Filomat \textbf{29} (2015), pp. 479--492.

\bibitem{Desz03} R. Deszcz, \emph{On Roter type manifolds}, 5th Conference on Geometry and Topology of Manifolds, April 27 - May 3, 2003, Krynica, Poland.

\bibitem{DPSch-2013} R. Deszcz, M. Plaue and M. Scherfner, %\emph{On Roter type warped products with 1-dimensional fibres},
J. Geom. Phys. \textbf{69} (2013) 1.

\bibitem{Glog-2007} M. G\l ogowska, %\emph{On Roter type manifolds},
Pure and Applied Differential Geometry-PADGE, (2007), pp. 114--122.


\bibitem{DGP-TV-2011} R. Deszcz, M. G\l ogowska, M. Petrovi\'c-Torga\u{s}ev and L. Verstraelen, %\emph{On the Roter type of Chen ideal submanifolds},
Results Math. \textbf{59} (2011) 401.


\bibitem{SB08} A. A. Shaikh and T. Q. Binh, %\emph{On some class of Riemannian manifolds,}
Bull. Transilv. Univ. \textbf{15(50)} 351 (2008).

\bibitem{SS06} A. A. Shaikh and S. K. Jana, %\emph{On weakly cyclic Ricci symmetric manifolds,}
Ann. Pol. Math. \textbf{89(3)} 139 (2006).

\bibitem{SS07} A. A. Shaikh and S. K. Jana, %\emph{On quasi-conformally flat weakly Ricci symmetric manifolds,}
Acta Math. Hungar. \textbf{115(3)} 197 (2007).

\bibitem{Gray78} A. Gray, %\emph{Einstein-like manifolds which are not Einstein},
Geom. Dedicata, \textbf{7} 259 (1978).

\bibitem{S81} U. Simon, \emph{Codazzi tensors}, Glob. Diff. Geom. and Glob. Ann. (Lecture notes, 838, Springer-Verlag, (1981), pp. 289--296.

\bibitem{F81} D. Ferus, \emph{A remark on Codazzi tensors on constant curvature space}, Glob. Diff. Geom. Glob. Ann. (Lecture notes 838, Springer, \textbf{1981}).

\bibitem{MM12a} C. A. Mantica and L. G. Molinari, %\emph{Extended Derdzinski-Shen theorem for curvature tensors},
Colloq. Math. \textbf{128}  1 (2012).

\bibitem{MM14} C. A. Mantica and L. G. Molinari, %\emph{Weyl compatible tensors},
Int. J. Geom. Meths. Mod. Phys. \textbf{11(08)}, 1450070 (2014).

\bibitem{DD91} F. Defever and R. Deszcz, %\emph{On semi-Riemannian manifolds satisfying the condition $R\cdot R=Q(S,R)$}, 
Geometry and Topology of Submanifolds III, World Sci., River Edge, NJ, (1991), pp. 108-130.

\bibitem{MM12b} C. A. Mantica and L. G. Molinari, %\emph{Riemann compatible tensors},
Colloq. Math. \textbf{128}  197 (2012).

\bibitem{SKP03} Y. J. Suh, J.-H. Kwon and Y. S. Pyo, %\emph{On semi-Riemannian manifolds satisfying the second Bianchi identity},
J. Korean Math. Soc. \textbf{40(1)} 129 (2003).

\bibitem{MS12a} C. A. Mantica and Y. J. Suh, %\emph{The closedness of some generalized curvature 2-forms on a Riemannian manifold I},
Publ. Math. Debrecen {\bf{81(3-4)}} 313 (2012) .

\bibitem{MS13a} C. A. Mantica and Y. J. Suh, %\emph{The closedness of some generalized curvature 2-forms on a Riemannian manifold II},
Publ. Math. Debrecen {\bf{82(1)}} 163 (2013).

\bibitem{MS14} C. A. Mantica and Y. J. Suh, %\emph{Recurrent conformal 2-forms on pseudo-Riemannian manifolds}, 
Int. J. Geom. Methods Mod. PhyS. \textbf{11(6)}, 1450056 (2014).


\bibitem{P95} M. Prvanovi$\acute{\mbox{c}}$, %\emph{On weakly symmetric Riemannian manifolds,}
Publ. Math. Debrecen \textbf{46(1-2)} 19 (1995).

\bibitem{Venz85} P. Venzi, %\emph{Una generalizzazione degli spazi ricorrenti},
Rev. Roumaine Math. Pures Appl. \textbf{30} 295 (1985).


%%%%%%%%%%%%%%%%%%%%%%%%%%%%%%%%%%%%%%%%%%%%%%%%%%%%%%%%%%%%%%%%%%%%%%%%%%%%%%%%%%%%%%%%%%%%%%%%%%%%%%%%%%

\end{thebibliography}
\end{document}